\documentclass{article}
\usepackage[utf8]{inputenc}
\usepackage{graphicx}
\usepackage{amsmath}
\usepackage{amssymb}
\usepackage{multirow}
\usepackage{color}
\usepackage{listings}
\usepackage{float}
\usepackage{authblk}
\usepackage{algorithm}
\usepackage{algpseudocode}

\newcommand{\eat}[1]{}

\usepackage[
  pass,
]{geometry}
\begin{document}

\title{Adaptive Cuckoo Filters}

\author[1]{Michael Mitzenmacher}
\author[2]{Salvatore Pontarelli}
\author[3]{Pedro Reviriego}
\affil[1]{ Harvard University, 33 Oxford Street, Cambridge, MA 02138, USA.\\
  \texttt{michaelm@eecs.harvard.edu}}
\affil[2]{Consorzio Nazionale Interuniversitario per le Telecomunicazioni (CNIT), Via del Politecnico 1, 00133 Rome, Italy.\\
  \texttt{salvatore.pontarelli@uniroma2.it}}
\affil[3]{Universidad Antonio de Nebrija, C/ Pirineos, 55, E-28040 Madrid, Spain.\\
  \texttt{previrie@nebrija.es}}
\maketitle

\begin{abstract}
We introduce the adaptive cuckoo filter (ACF), a data structure for approximate set membership that extends cuckoo filters by reacting to false positives, removing them for future queries.  As an example application, in packet processing queries may correspond to flow identifiers, so a search for an element is likely to be followed by repeated searches for that element. Removing false positives can therefore significantly lower the false positive rate.  The ACF, like the cuckoo filter, uses a cuckoo hash table to store fingerprints.  We allow fingerprint entries to be changed in response to a false positive in a manner designed to minimize the effect on the performance of the filter. We show that the ACF is able to significantly reduce the false positive rate by presenting both a theoretical model for the false positive rate and simulations using both synthetic data sets and real packet traces.
\end{abstract}

\section{Introduction}

Modern networks need to classify packets at high speed. A common function in packet processing is to check if a packet belongs to a set $S$.  As a simple example, if the source of the packet is on a restricted list, the packet should be subject to further analysis. In some applications, the cost of such a check can be beneficially reduced via an approximate membership check that may produce false positives but not false negatives;  that is, it may with small probability say an element is in a set even when it is not.  We first use the approximate membership check, and if we are told the element is not in the set we can safely continue regular processing.  On a positive response a more expensive full check may be done to find the false positives, removing them from further analysis.  Packets with a source on the restricted list are always correctly identified.  This approach speeds the common case by storing the approximate representation of the set $S$ in a small fast memory, and the exact representation of $S$ in a bigger but slower memory that is accessed infrequently.  

The well known example of a data structure for approximate set membership is the Bloom filter \cite{Bloom}, which is widely used in networking applications. Many variants and enhancements for Bloom filters have been proposed \cite{BloomApp}. There are other data structures that also provide approximate membership checks with improved performance, such as the recently proposed cuckoo filter \cite{CF} that stores fingerprints of the elements in a suitable cuckoo table \cite{CH}, as we describe in more detail in Section 2.  A key parameter for such data structures is the false positive probability, which gives the probability that an element not in the set yields a false positive.  A slightly different quantity is the false positive rate, which gives the fraction of false positives for a given collection of queries (or, in some settings, for a given distribution of the input queries).  If an element that gives a false positive is queried for many times, the false positive rate may be much higher than the false positive probability, even though in expectation over suitably randomly chosen hash functions the false positive rate should be close to the false positive probability.  

In many network applications, the queried elements are repeated. For example, let us consider an application in which we track the packets of a subset of the 5-tuple flows on a link. We may use an approximate membership check for each packet to tell us if the packet belongs to one of the flows that we are tracking, in which case it is subject to more expensive processing.  In this application, receiving a packet with 5-tuple $x$ makes it very likely that we will receive other packets with the same 5-tuple value in the near future. Now suppose that a packet with 5-tuple $x$ returns a false positive on the approximate membership check. It is clear that we would like to adapt the approximate membership check structure so that $x$ will not return a false positive for the following packets of the same flow.  Ideally, the false positive responses could theoretically be reduced to one per flow, lowering the false positive rate. 

Such adaptation cannot be easily done for the existing approximate membership check structures. A variation of a Bloom filter, referred to as a retouched Bloom filter, removes false positives (by changing some bits in the filter to 0), but can create false negatives by doing so \cite{RTBF}.  Other options include changing the number of hash functions according to the likelihood of an element being in a set \cite{PCBF},\cite{WBF}, but this approach requires additional offline information and computation that does not appear natural for many settings. In all of these cases, removing the false positives significantly changes the nature of the data structure. 

We introduce the adaptive cuckoo filter (ACF), an approximate membership check structure that is able to selectively remove false positives {\em without} introducing false negatives.  The scheme is based on cuckoo hashing \cite{CH} and uses fingerprints, following standard cuckoo filters \cite{CF}. In contrast with the cuckoo filter, where movements in the cuckoo table can be based solely on the fingerprints, here we assume we have the original elements available, although they can be in a slower, larger memory.  When inserting a new element, existing elements can be moved.  Our enhancement is to allow different hash functions for the fingerprints, which allows us to remove and reinsert elements to remove false positives without affecting the functionality of the filter. The removal of false positives is almost as simple as a search operation, and does not substantially impact performance.  Instead, the impact is felt by having a slightly more complex insertion procedure, and requiring slightly more space than a standard cuckoo filter to achieve the same false positive probability.  Of course, since the ACF is adaptive, it generally achieves a better false positive rate than a standard cuckoo filter with the same space, and our goal here is to improve the false positive rate (as opposed to the false positive probability).  

We provide a theoretical model that provides accurate estimates for ACF performance.  We validate ACF performance through simulations using synthetic data and real packet traces.  The results confirm that the ACF is able to reduce false positives, making it of interest for many networking applications. 

Before beginning, we emphasize that the ACF is not a general replacement for a Bloom filter or cuckoo filter, but an optimization that is particularly well-suited to the setting when a pre-filter is desired to avoid unnecessary accesses to memory for data that is not stored there.  There are many applications where the original, complete data is required in any case, and the filter is used as a first stage screen to prevent the need to access slower memory to check the original data.  For example, for both whitelists and blacklists, a pre-filter could make lookups much more efficient, but all proposed positives would be checked against the original data.  Even if only an approximate set membership data structure without the original data is required, in some architectures the memory cost of a slower and larger memory to hold the original data may be relatively inexpensive.  However, we note that our need for storing the original data might make the use of ACFs unsuitable for many applications.   

The rest of the paper is organized as follows.  Section 2 briefly reviews the data structures on which the ACF is based, namely cuckoo hash tables and cuckoo filters. Section 3 introduces the ACF, describing its implementation and providing a model of its expected performance. The ACF is evaluated in Section 4 using both randomly generated packets and real packet traces to show its effectiveness. We conclude with a summary and some ideas for future work.  

\section{Preliminaries}

We provide a brief overview of cuckoo hashing \cite{CH} and cuckoo filters \cite{CF}.

\subsection{Cuckoo hashing}

Cuckoo hash tables provide an efficient dictionary data structure \cite{CH}. A cuckoo hash table can be implemented as a collection of $d$ subtables composed of $b$ buckets each, with each bucket having $c$ cells. The hash table uses $d$ hash functions $h_1,h_2,\ldots,h_d$, where the domain is the universe of possible input elements and the range is $[0,b)$. We assume that all functions are independent and uniform;  such assumptions are often reasonable in practice \cite{MV}.  A given element $x$ can be placed in bucket $h_i(x)$ in the $i$th subtable;  each element is placed in only one bucket.  The placement is done so that there are at most $c$ elements in any bucket.  This limitation can require moving elements among their choices of buckets when a new element is inserted.  Values associated with elements can be stored with them if that is required by the application;  alternatively, pointers to associated values kept in an external memory can also be stored.   
The structure supports the following operations: \\


{\bf Lookup:} Given an element $x$, the buckets given by the $h_i(x)$ are examined to see if $x$ is in the table.

{\bf Insertion:} Given an element $x$, we first check that $x$ is not already in the table via a lookup.  If not, we sequentially examine the buckets $h_i(x)$.  If there is empty space, $x$ is inserted into the first available space. If no space is found, a value $j$ is chosen independently and uniformly at random from $[1,d]$, and $x$ is stored in the $j$th subtable at bucket $h_j(x)$. This displaces an element $y$ from that bucket, and then the insertion process is recursively executed for $y$. 

{\bf Deletion:} Given an element $x$, $x$ is searched for via a lookup; if it is found it is removed.\\

Cuckoo hash tables are governed by a threshold effect;  asymptotically, if the load, which is the number of elements divided by the number of cells, is below a certain threshold (which depends on $c$ and $d$), then with high probability all elements can be placed.

The above description of insertion is somewhat incomplete, as there is room for variation.  For example, the displaced element $y$ is often chosen uniformly at random from the bucket, and it is often not allowed to put $y$ immediately back into the same bucket.  The description also does not suggest what to do in case of a failure.  Generally after some number of recursive placement attempts, a failure occurs, in which case elements can be re-hashed, or an additional structure referred to as a stash can be used to hold a small number of items that would not otherwise be placed \cite{stash}.
  
An alternative implementation of cuckoo hashing uses a single table of buckets, so each hash function can return any of the buckets. The two alternatives provide the same asymptotic performance in terms of memory occupancy.  

As previously stated, the asymptotic load threshold depends on the values of $c$ and $d$, the number of cells per bucket and the number of hash functions.  In practice, even for reasonably sized systems one obtains performance close to the asymptotic load threshold.  Two natural configurations include $d=2$ and $c=4$, which gives a threshold over 98\%, and $d=4$ and $c=1$ which gives a threshold of over 97\%.  These configurations give high memory utilization, while requiring a small memory bandwidth \cite{CHocc}. The decision for what configuration to use may depend on the size of the item being stored and the natural cell size for memory.  

Note that if subtables can be put on distinct memory devices, then lookup operations can possibly be completed in one memory access cycle by searching subtables in parallel \cite{PCH}.

\subsection{Cuckoo filters}

A cuckoo filter is an approximate membership check data structure based on cuckoo hash tables \cite{CF}.  Instead of storing set elements, a cuckoo filter stores fingerprints of the elements, obtained using an additional hash function.  In this way, it uses less space, while still being able to ensure a low false positive probability.  In the suggested implementation, each element is hashed to $d=2$ possible buckets, with up to $c=4$ elements per bucket. 

To achieve small space, the cuckoo filter does not store the original elements.  A difficulty arises in moving the fingerprints when buckets are full, as is required by the underlying cuckoo hash table.  The cuckoo filter uses the following method, referred to as partial-key cuckoo hashing.  If the fingerprint of an element $x$ is $f(x)$, its two bucket locations will be given by $h_1(x)$ and $h_1(x) \oplus h_2(f(x))$.  Notice that, given a bucket location and a fingerprint, the other bucket location can be determined by computing $h_2(f(x))$ and xoring the result with the current bucket number.  

The structure supports the following operations, where in what follows we assume an insertion is never performed for an element already in the structure: \\

{\bf Lookup:}  Given an element $x$, compute its fingerprint $f(x)$, and the bucket locations $h_1(x)$ and $h_1(x) \oplus h_2(f(x))$. The fingerprints stored in these buckets are compared with $f(x)$.  If any fingerprints match $f(x)$, then a positive result is returned, otherwise a negative result is returned. \

{\bf Insertion:}  Given an element $x$, compute the two bucket locations and the fingerprint $f(x)$.  If a cell is open in one of these buckets, place $f(x)$ in the first available open cell.  Otherwise, a fingerprint $f(y)$ in one of the buckets is displaced, and then recursively placed, in the same manner as in a cuckoo hash table.  Here we use that both buckets for $y$ can be determined from $f(y)$ and the bucket from which $f(y)$ was displaced. 

{\bf Deletion:} Given an element $x$, $x$ is searched for via a lookup. If the fingerprint $f(x)$ is found in one of the buckets, one copy of the fingerprint is removed.\\ 

The false positive probability of a cuckoo filter can be roughly estimated as follows. If $a$ bits are used for the fingerprints and the load of the hash table is $\ell$, the false positive probability will be approximately $8\ell /2^{a}$. This is because on average the search will compare against $8\ell$ fingerprints (assuming 2 choices of buckets with 4 cells per bucket), with each having a probability $2^{-a}$ of yielding a false positive. 

\section{Adaptive Cuckoo Filter Construction}

Before beginning, we note that pseudocode for our algorithms appear in the Appendix.    

Our proposed ACF stores the elements of a set $S$ in a cuckoo hash table.  A replica of the cuckoo hash table that stores fingerprints instead of full elements is constructed, and acts as a cuckoo filter. The key difference from the cuckoo filter is that we do {\em not} use partial-key cuckoo hashing;  the buckets an element can be placed in are determined by hash values of the element, and not solely on the fingerprint. To be clear, the filter uses the same hash functions as the main cuckoo hash table;  the element and the fingerprint are always placed in corresponding locations.   Using the filter, false positives occur when an element not in the set has the same fingerprint as an element stored in one of the positions accessed during the search.  A false positive is detected by examining the cuckoo hash table when a positive result is found;  if the element is not found in the corresponding location -- specifically, in the same bucket and same position within that bucket as the corresponding fingerprint -- a false positive has occurred, and moreover we know what element has caused the false positive.  As we explain below, to remove the false positive, we need to change the fingerprint associated with that element, by using a different fingerprint function.  We suggest two ways of accomplishing this easily below.    

Before describing the structure in more detail, we define the parameters that will be used.
\begin{itemize}
\item The number of tables used in the Cuckoo hash: $d$.
\item The number of cells per bucket: $c$.
\item The total number of cells over all tables: $m$.
\item The number of buckets per table: $b = m/(d \cdot c)$.
\item The occupancy or load of the ACF: $\ell$.
\item The number of bits used for the fingerprints: $a$.
\end{itemize}

To see the potential benefits of the ACF, let us consider its behavior over a window of time.  Let us suppose that a fraction $p_1$ of operations are lookups, $p_2$ are insertions, and $p_3$ are deletions.  Let us further suppose that of the lookup operations, a fraction $q_1$ are true positives, and the remaining fraction $1-q_1$ have a false positive rate of $q_2$.  If we assume costs $c_I,c_D,c_P,$ and $c_F$ for the inserts, deletes, positive lookups, and negative lookups, we find the total cost is:
$$p_1((q_1 +(1-q_1)q_2)c_P + (1-q_1)(1-q_2)c_F) + p_2c_I + p_3 c_D.$$  
In the types of implementations we target, we expect $p_1$ to be large compared to $p_2$ and $p_3$, so that the cost of lookups dominate the costs of insertions and deletions.  We also expect $q_1$ to be small, and $c_F$ to be much less than $c_P$, so a filter can successfully and at low cost prevent lookups to slow memory.  In this case the dominant cost corresponds to the term $p_1(q_1 +(1-q_1)q_2)c_P$, and as $q_1$ is small, the importance of minimizing the false positive rate $q_2$ is apparent.  

We now present variants of the ACF and discuss their implementation.

\subsection{ACF for buckets with one cell}

We present the design of the ACF when the number of tables is $d=4$ and the number of cells per bucket is $c=1$. This configuration achieves similar utilization as $d = 2$ and $c = 4$ and requires less memory bandwidth \cite{CHocc} at the cost of using more tables. 
The ACF consists of a filter and a cuckoo hash table, with both having the same number of tables and numbers of cell per bucket, so that there is a one-to-one correspondence of cells between the two structures. The ACF stores only a fingerprint of the element stored in the main table.

To implement the ACF we use a small number of bits within each bucket in the ACF filter to represent a choice of hash function for the fingerprint, allowing multiple possible fingerprints for the same element. We use $s$ hash selector bits, which we denote by $\alpha$, to determine which hash function is used to compute the fingerprint stored in that bucket. We refer to the fingerprint hash functions as $f_0,f_1,\ldots,f_{2^s-1}$.  For example, by default we can set $\alpha=0$ and use $f_0(x)$. If we detect a false positive for an element in that bucket then we increment $\alpha$ to 1 and use $f_1(x)$ (For convenience we think of the $s$ bits as representing values 0 to $2^s-1$). 
In most cases this will eliminate the false positive. On a search, we first read the hash selector bits, and then we compute the fingerprint using the corresponding fingerprint function before comparing it with the stored value. 
The downside of this approach is we are now using a small number of additional bits per cell to represent the choice of hash function, increasing the overall space required.  

In more detail, to remove a false positive, we increment $\alpha$ (modulo $2^s$) and update the fingerprint on the filter accordingly (Sequential selection is slightly better than random selection as it avoids picking a value of $\alpha$ that recently produced another false positive). We verify that the adaptation removes the false positive by checking that the new fingerprint is different from the new fingerprint for the element that led to the false positive. For example, if $x$ created a false positive with an element $y$ stored on the ACF when $\alpha = 1$, we can increment $\alpha$ and check that $f_2(x) \neq f_2(y)$;  if the fingerprints match we can increment $\alpha$ again.  This refinement is not considered in the rest of the paper as the probability of removing the false positive is already close to one $(1-1/2^{a})$ and we want to keep the adaptation procedure as simple as possible. 


The insertion and deletion procedures use the standard insertion or deletion methods for the cuckoo hash table, with the addition that any insertions, deletions, or movement of elements are also performed in the filter to maintain the one-to-one relationship between elements and fingerprints.

To perform a search for an element $x$, the filter is accessed and buckets $h_1(x),\ldots,h_4(x)$ are read. From each bucket the corresponding fingerprint and $\alpha$ value is read and the fingerprint value is compared against $f_{\alpha}(x)$. If there is no match in any table, the search returns that the element is not found.  If there is a match, the cuckoo hash table is accessed to see if $x$ is indeed stored there. If there is a match but $x$ is not found, a false positive is detected. The search in the filter of an ACF is similar to a search in a cuckoo filter and should have a similar false positive probability;  however, we can reduce the false positive rate.  We emphasize again that the false positive probability corresponds to the probability that a ``fresh,'' previously unseen element not in the set yields a false positive. Instead, the false positive rate corresponds to the fraction of false positives over a given sequence of queries, which may repeatedly query the same element multiple times.  

To model the behavior of the ACF, we will define a Markov chain that describes the evolution of $\alpha$ in a single bucket of the filter, with changes to $\alpha$ triggered by the occurrence of false positives. We assume that we have $\zeta$ elements that are not in the hash table that are currently checked on a given bucket; this is the set of potential false positives on that bucket. We assume that requests from the $\zeta$ elements are generated independently and uniformly at random; while arguably this assumption is not suitable in many practical situations, it simplifies the analysis, and is a natural test case to determine the potential gains of this approach.  (More skewed request distributions will generally lead to even better performance for adaptive schemes.) We further assume that the hash table is stable, that is no elements are inserted and deleted, in the course of this analysis of the false positive rate.  Finally, we first perform the analysis from the point of view of a single bucket;  we then use this to determine the overall expected false positive rate over all buckets. We analyze the false positive rate assuming that these are the only queries; true matches into the hash table are not counted in our analysis, as the frequency with which items in the table are queried as compared to items not in the table is application-dependent.

Without loss of generality we may assume that an element $x$ is in the
hash table, and the fingerprint value $f_0(x)$ is located in the cell.  (If there
is no element in the cell, then there are no false positives on that cell.)  The
number of elements that provide a false positive on the first hash
function is a binomial random variable $\mbox{Bin}(\zeta,2^{-a})$, which
we approximate by a Poisson random variable with mean
$\zeta2^{-a}$ for convenience.  We refer to this value by the random
variable $Z_1$, and similarly use $Z_j$ to refer to the corresponding
number of elements that provide a false positive for the $j$-th hash
function.  It is a reasonable approximation to treat the $Z_j$ as
independent for practical values of $\zeta$ and $a$, and we do so henceforth.    

We can now consider a Markov chain with states $0,1,\ldots,2^s-1$; the
state $i$ corresponds to $\alpha = i$, so the cell's fingerprint is $f_i(x)$.  
The transitions of this Markov chain are from $i$ to $i+1$ modulo $2^s$ with probability
$Z_i/\zeta$; all other transitions are self-loops. Note that if any of
the $Z_i$ are equal to zero, we obtain a finite number of false
positives before finding a configuration with no additional false
positives for this set of $\zeta$ elements; if all $Z_i$ are greater than
zero, then we obtain false positives that average to a constant rate
over time.  Using this chain, we can determine the false positive
rate.

Specifically, let us suppose that we run for $n$ queries.  Let $\mu
= \zeta 2^{-a}$, so that $e^{-\mu}$ is our approximation for the
probability that there are no false positives for a given hash
function.  We first note that for any value $k < 2^s$ we have a probability 
$$P_{stop}(k)=(1-e^{-\mu})^k e^{-\mu}$$
to finish in a configuration with no additional false positives after the bucket is triggered by $k$ false positives. For this case the false positive rate is simply $k/n$, if we suppose that $n$ is sufficiently large to trigger the system up to the final state.

For the case where all hash functions have at least one element that
yields a false positive, so that all the $Z_i$ are greater than 0, 
the expected number of queries to complete
a cycle from the first hash function and back again, given the $Z_i$ values, is given by the 
following expression:
$$\sum_{j=0}^{2^s-1} \frac{\zeta}{Z_j}.$$
Over this many queries, we obtain $2^s$ false positives, 


The overall asymptotic false positive rate $F(\zeta,n)$ for a bucket which stores an element and to which $\zeta >0$ elements not in the hash table map can therefore be expressed as:


$$F(\zeta,n) = \sum_{i=1}^{2^s-1} (1-e^{-\mu})^i e^{-\mu} \frac{i}{n} + \sum_{Z_0,Z_1,\ldots,Z_{2^s-1} > 0} \left ( \prod_{i=0}^{2^s-1} \frac{e^{-\mu} \mu^{Z_i}}{Z_i!} \right )  \frac{2^s}{\sum_{i=0}^{2^s-1} \frac{\zeta}{Z_i}}.$$

Notice that this expression is a good approximation even for reasonably small values of $n$.  
Further, when $n$ is large, the first term will be comparatively small compared to the second. 

The previous analysis considered a bucket in isolation.  We now move to consider the false positive rate for an entire table that has $b d$ buckets with an occupancy $\ell$.  The number of elements $\zeta$ that map to each bucket is a binomial random variable, asymptotically well-approximated by a Poisson random variable. Let us assume that there are $A$ elements not stored that are searched for in the table; then $\zeta$ can be approximated by a Poisson random variable with mean $e_b = A/b$.  Over a collection of $N$ queries to the table, the total number of queries $j$ that map to a bucket with $\zeta$ items is itself a binomial random variable, asymptotically well-approximated by a Poisson random variable with mean $N\zeta/A$.  We conclude that the (asymptotic, approximate) average false positive rate $F$ over all buckets is:



$$ F = b d \ell \sum_{\zeta > 0} \frac{e^{-e_b}{e_b}^\zeta}{\zeta!} \sum_{j \geq 0} \frac{e^{-\zeta N/A} {(\zeta N/A)}^j F(\zeta,j) (j/N)}{j!}.$$

While we do not provide a full proof here, we note that standard concentration results can be used to show that in this model the false positive rate over all the buckets is close to the expected false positive rate calculated above.  Such concentration means we should not see large variance in performance over instantiations of the ACF structure in this setting.  

Although we have looked at a model where each query is equally likely to come from each of the $\zeta$ flows, so each has approximately the same size $n_e$, we could generalize this model to settings where the $\zeta$ flows correspond to a small number of types (or rates), and then consider similar calculations based on the number of false positives of each rate.  Intuitively, the setting we have chosen where each flow is equally likely to be queried at each step is the worst case for us, as it leads to the fastest cycling through the $2^s$ fingerprint hash functions in the case that each fingerprint has at least one false positive. This will be confirmed by the results presented in the evaluation section.   

\subsection{ACF for buckets with multiple cells}

We now examine the case where the number of tables is $d=2$ and the number of cells per bucket is $c=4$, the case that was studied originally in constructing cuckoo filters \cite{CF}. The ACF again consists of a filter and a cuckoo hash table with a one-to-one correspondence of cells between the two structures; the ACF holds the fingerprint of the element stored in the main table.  To reduce the false positive rate, the hash functions used for the fingerprint are different for each of the cell locations, so that in our example there will be four possible fingerprints for an element, $f_{1}(x)$, $f_{2}(x)$, $f_{3}(x)$ and $f_{4}(x)$, each corresponding to a cell (1, 2, 3, and 4) in the bucket. 

\eat{
The overall structure is illustrated in Figure \ref{fig:1}. 
\begin{figure}[t]
	\includegraphics[width=0.5\textwidth]{figs/Fig1.pdf}
	\caption{Structure of the ACF with $d$ = 2 and $c$ = 4.}
	\label{fig:1}
\end{figure}
}

The insertion and deletion procedures use the standard insertion or deletion methods for the cuckoo hash table, with the addition that any changes are matched in the filter to maintain the one-to-one relationship between elements and fingerprints.  Note that if an element is moved via insertion to a different cell location, its fingerprint may correspondingly change;  if an element $x$ moves from cell 1 to cell 2 (either in the same bucket or a different bucket), its fingerprint changes from $f_1(x)$ to $f_2(x)$.    

The search is similar to the case described in the previous subsection. When there is a match in the filter but not the main table, a false positive is detected. 

We describe how this ACF responds when a false positive occurs. Suppose a search for an element $x$ yields a false positive; the fingerprints match, but element $y$ is stored at that location in the cuckoo hash table.  For convenience let us assume that the false positive was caused by the fingerprint stored in the first cell of the first table, and that there is another element $z$ in the second cell on the first table. Then, we can swap $y$ and $z$ in the ACF, keeping them in the same bucket, so that now instead of $f_{1}(y)$ and $f_{2}(z)$, the bucket will hold $f_{1}(z)$ and $f_{2}(y)$ in the first two cells. If subsequently we search again for $x$, in most cases $f_{1}(z)$ and $f_{2}(y)$ will be different from $f_{1}(x)$ and $f_{2}(x)$, and therefore the false positive for $x$ will be removed. If we search for $y$ or $z$, we still obtain a match. As a result of the change, however, we might create false positives for other elements that have also been recently queried.  
More precisely, when a lookup provides a positive response but the cuckoo hash table does not have the corresponding element, there was a false positive and the filter adapts to remove it by randomly selecting one of the elements in the other $c-1$ cells on the bucket and performing a swap. 

To model the ACF for buckets with multiple cells, we define a Markov chain that 
takes into account the evolution of a bucket as cells are swapped.  
As before we assume fingerprints of $a$ bits, with $\zeta$ elements creating
potential false positives, and each request being independently and
uniformly chosen from this set of elements. Without loss of generality we
may assume that elements $w,x,y,$ and $z$ are in the bucket being
analyzed, initially in positions $1,2,3,$ and $4$ respectively.  Let
$Z_{w,1}$ be the number of elements that yield a false positive with element
$w$ when it is in position 1, that is when the corresponding value is
$f_1(w)$, and similarly for the other elements and positions.  
Here, since some buckets may not be full, we can think of some of these elements as 
possibly being ``null'' elements that do not match with any of the $\zeta$
potential false positives.  For non-null elements, as before,
$Z_{w,1}$ is distributed as a binomial random variable
$\mbox{Bin}(\zeta,2^{-a})$, approximated by Poisson random variable
with mean $\mu = \zeta2^{-a}$, and we treat the random variables as
independent.  We use
$\hat{Z}$ to refer to a 16-dimensional vector of values for $Z_{w,1}$, $Z_{w,2}$, etc.

In this setting, we consider a Markov chain with states being the
24 ordered tuples corresponding to all possible orderings of $w,x,y,$
and $z$.  The transition probability from for example the ordering
$(w,x,y,z)$ to $(x,w,y,z)$ is given by
$$\frac{Z_{w,1}}{3\zeta} + \frac{Z_{x,2}}{3\zeta};$$ that is, with
probability $Z_{w,1}/\zeta$ the query is for an element that gives
a false positive against $w$, and then with probability $1/3$ the
elements $x$ and $w$ are swapped, and similarly for the second term
in the expression.  Any transition corresponds to either a swap of two elements or
a self-loop. It is possible an absorbing state can be reached where there is no
transition that leaves that state -- that is, there are no false positives in the given state
-- if there is some permutation where the corresponding $Z$ variables for transitions
out of that state
are all 0.  However, there may be no such absorbing state, in which case the 
false positives average to a constant rate over time.  (This is similar to what
we have previously seen in the ACF with one cell per bucket.)  While there is no simple way to write an expression for the false positive rate for a bucket based on $\hat{Z}$, as 
there was in the setting of one cell per bucket, the calculations are straightforward.

Given values for a vector of $\hat{Z}$, we can determine the
false positive rate by determining either the expected number of 
transitions to reach the stable state, or the expected overall
rate of false positives, depending on the values.  Thus, in theory, we can determine the overall expected false positive rate, similarly to the case of buckets with
one cell.  That is, we can calculate the probability of each vector $\hat{Z}$,
and determine either the probability distribution on the number of 
false positives before reaching an absorbing state (if one exists), or 
the expected time between false positives (which can be derived from 
the equilibrium distribution if there is no absorbing state), and take the corresponding weighted sum. In practice we would sum over vectors $\hat{Z}$ of sufficient probability to obtain an approximation good to a suitable error tolerance; however, we find the number of relevant vectors $\hat{Z}$ is very large, making this computation impractical. 

Instead, we suggest a sampling-based approach to determine the false positive rate.  Vectors $\hat{Z}$ are sampled a large number of times with the appropriate probability, with the corresponding false positive rate for a bucket determined for each sampled $\hat{Z}$, to obtain a reasonable approximation. This sampling process must take into account that some cells may be left unoccupied. Based on
this, we can calculate an estimate for the expected false positive rate for a single bucket to which $\zeta$ elements map, and then find the overall false positive rate by taking appropriate weighted averages, again similarly to the case of one cell per bucket.  



\subsection{Implementation considerations}

Both variants of the ACF we have proposed are simple, and straightforward to implement.
The search operation is similar to that of a cuckoo hash table, with the cuckoo filter acting as a prefilter.  Indeed, if there is no false positive, the cuckoo filter will determine what bucket the element is in, so that only one bucket in the cuckoo hash table will need to be checked for the element.  On a false positive, after checking the table the additional work to modify the table to adapt requires only a few operations.  

Both variants of the ACF require the computation of different hash functions to obtain the fingerprints. This adds some complexity to the implementation. However the overhead is expected to be small as hash functions can be computed in very few clock cycles (e.g. CLHASH is able to process 64 bytes in one clock cycle \cite{hashimp1}) in modern processors exploiting the last x86 ISA extensions \cite{hashimp2}  or implemented in hardware with low cost. In both variants, it may be possible to compute multiple fingerprints of an element at once, since most hash functions return either 32 or 64 bits, and these bits can be split into multiple fingerprints. In a hardware or modern processor implementation, hash computations are commonly much faster than memory accesses. Therefore, the time required to perform a lookup in the filter is typically dominated by the memory accesses to the filter cells. The number of such memory accesses is the same for both the cuckoo filter and the ACF. Therefore, we can argue that they will achieve similar lookup speeds.


\eat{
\begin{figure*}[t]
	\includegraphics[width=0.5\textwidth]{figs/Fig3-left.pdf}
    	\includegraphics[width=0.5\textwidth]{figs/Fig3-right.pdf}
	\caption{Search operations in ACF, $c$ > 1 left and $c$ = 1 right.}
	\label{fig:1}
\end{figure*}
}

We have assumed thus far that adaptation, or the procedure to remove a false positive, is done every time a false positive is found.  This is straightforward, but might not be optimal in some settings, depending on the costs of changing the ACF cuckoo hash table and filter.  We leave consideration of alternative schemes that do not try to modify the ACF on each false positive for future work.
For our algorithms, we have not considered the additional possibility of moving items to another bucket when false positives are detected, although this is possible with cuckoo hashing.  Our motivation is in part theoretical and in part practical.  On the practical side, such movements would potentially be expensive, similar to additional cuckoo hash table insertions.  On the theoretical side, analyzing the false positive rate when items are moving in the table seems significantly more difficult. Again, considerations of such schemes would be interesting for future work.

\section{Evaluation}

To evaluate the performance of our proposed ACF implementations, we have simulated both with queries that are generated according to different parameters and also with real packet traces. In the first case, the goal is to gain understanding about how the performance of the ACF depends on the different parameters while in the second, the goal is to show that the ACF can provide reductions on the false positive rate for real applications. The results are also compared with those of the analytic estimates presented in Section 3. 

We note that we do not compare with possible alternatives such as the retouched Bloom filter or varying the number of hash functions because they are not directly comparable with the ACF; the retouched Bloom filter introduces false negatives, and varying the number of hash functions requires additional offline information.  The original cuckoo filter is the best natural alternative with which to compare.  

\subsection{Simulations with generated queries}

The first set of simulations aims to determine the effect of the different parameters on the performance of the ACF.
The ACF is filled with randomly selected elements up to 95\% occupancy; let $S$ be the number of elements stored in the ACF. Then $A$ elements that are not in the ACF are randomly generated.The queries are then generated by taking randomly elements from $A$ with all elements having the same probability and the false positive rate is measured. We consider the following configurations:

\begin{itemize}
\item ACF with $d = 4, c = 1$ (first variant) and $s = 1,2,3$.
\item ACF with $d = 2, c = 4$ (second variant).
\item Cuckoo filter with $d = 4, c = 1$.
\end{itemize}

We note that we only provide results for the cuckoo filter with $d = 4, c = 1$ and not for the cuckoo filter with $d=2, c=4$ in our comparisons as the former achieves a significantly better false positive rate than the latter.  To provide an initial understanding of the ACF performance, we fix the same probability of being selected for a query for all elements, so that on average each has $n_e$ queries. 
For the first variant we set the number of buckets to 32768 for each table, so the overall number of cells is 131072. For the second variant  there are two tables and each table has 16384 buckets. The overall number of cells is 131072 also in this case.
The number of bits for the fingerprint $a$ is set to 8,12 and 16 so that the standard cuckoo filter with $d = 4, c = 1$ gets a false positive rate of approximately 1.5\%, 0.1\%, and 0.006\% at 95\% occupancy. The same number of bits per cell is used for the different ACF configurations. This means that for the first variant the number of fingerprint bits are $8-s, 12-s$, and $16-s$. The ACFs are adapted on every false positive. The value of $n_e$ is set to 10, 100, and 1000 and the experiment is repeated for several ratios of elements not in the set ($A$) to elements in the set ($S$)  that appear in the queries ranging from 1 to 100. The $A/S$ ratios used in the experiment are $\{1,2,3,4,5,10,20,30,40,50,60,70,80,90,100\}$.

\begin{figure*}[ht]

\includegraphics[width=0.32\textwidth]{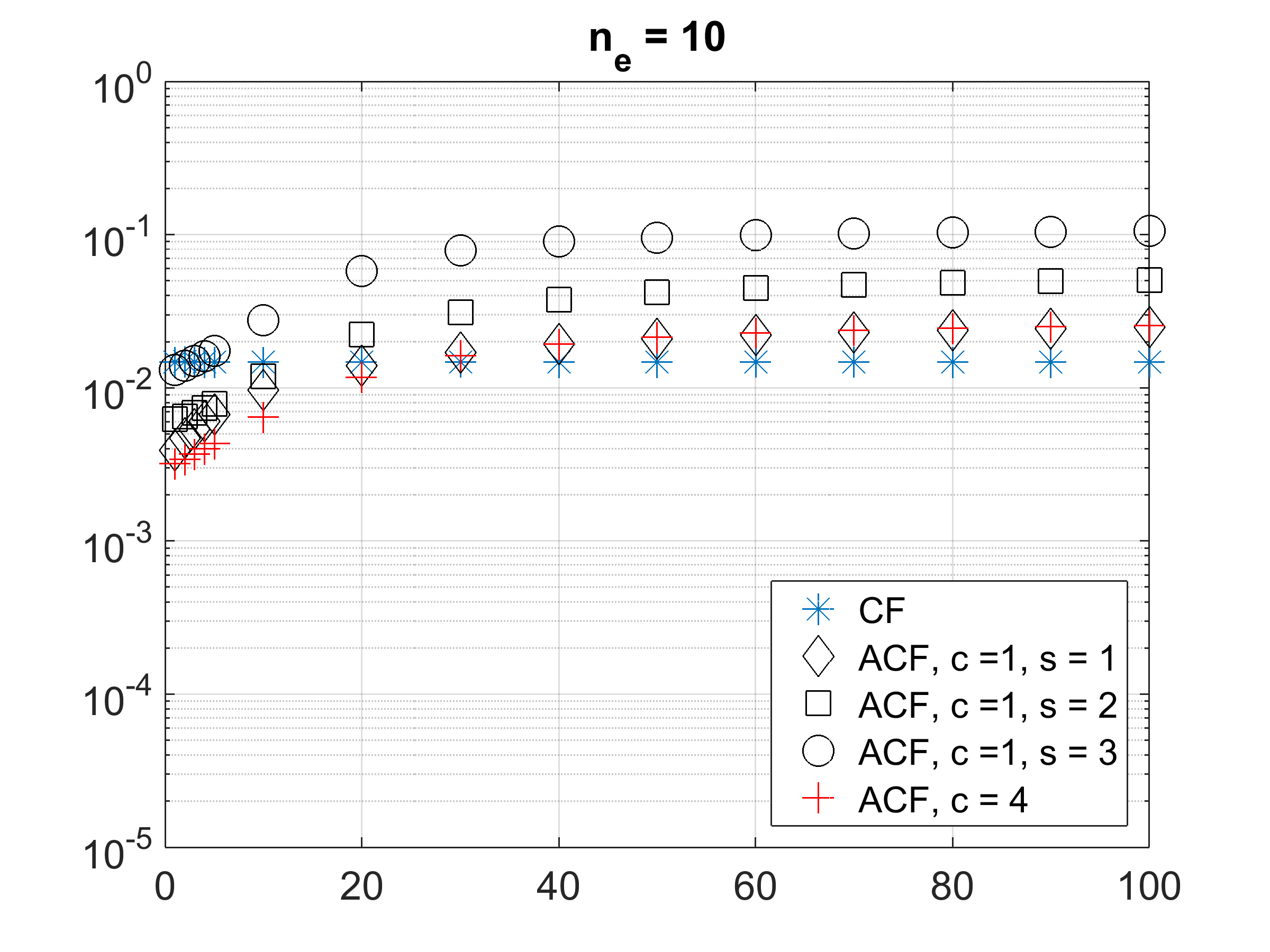}
\includegraphics[width=0.32\textwidth]{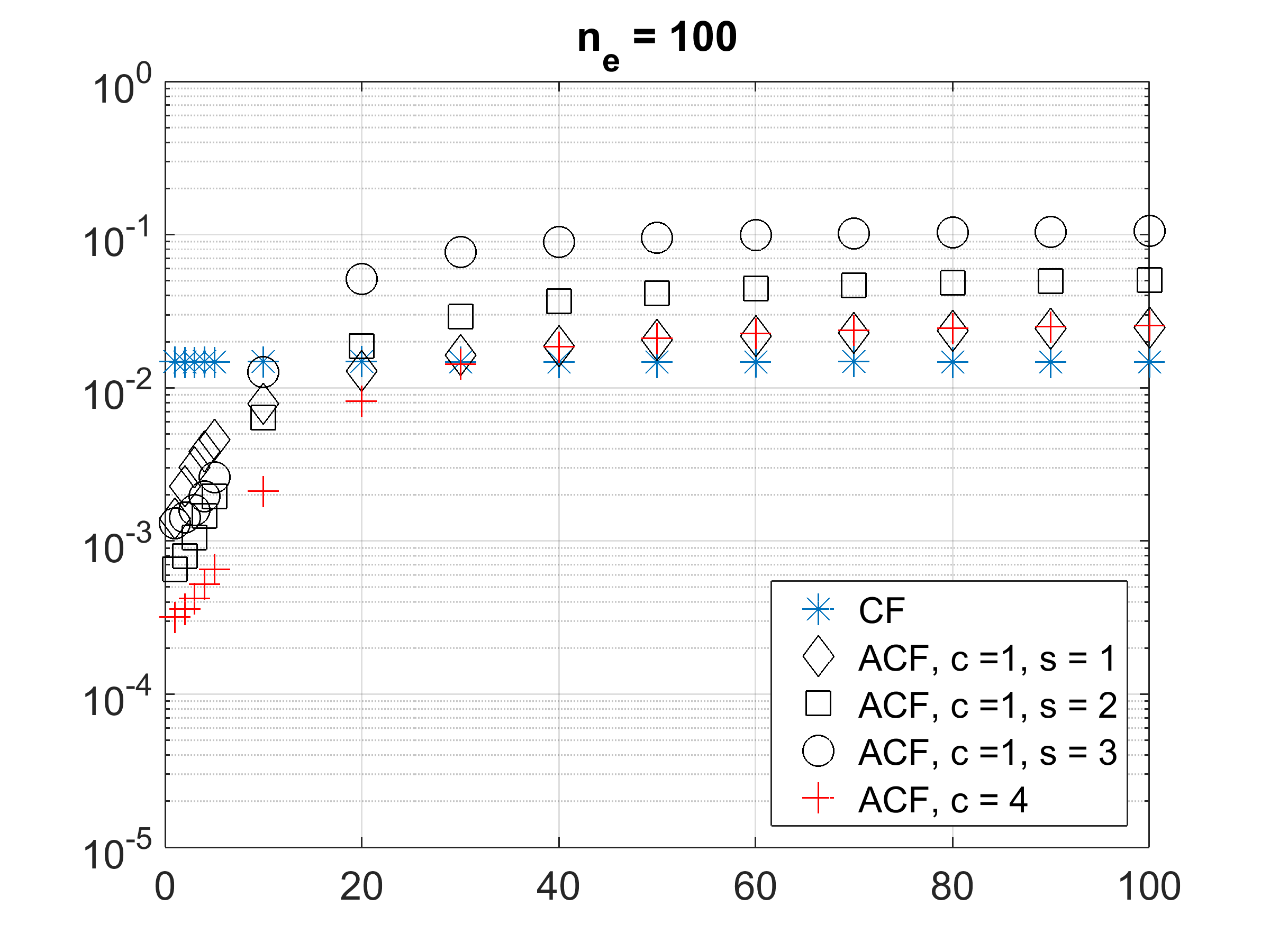}
\includegraphics[width=0.32\textwidth]{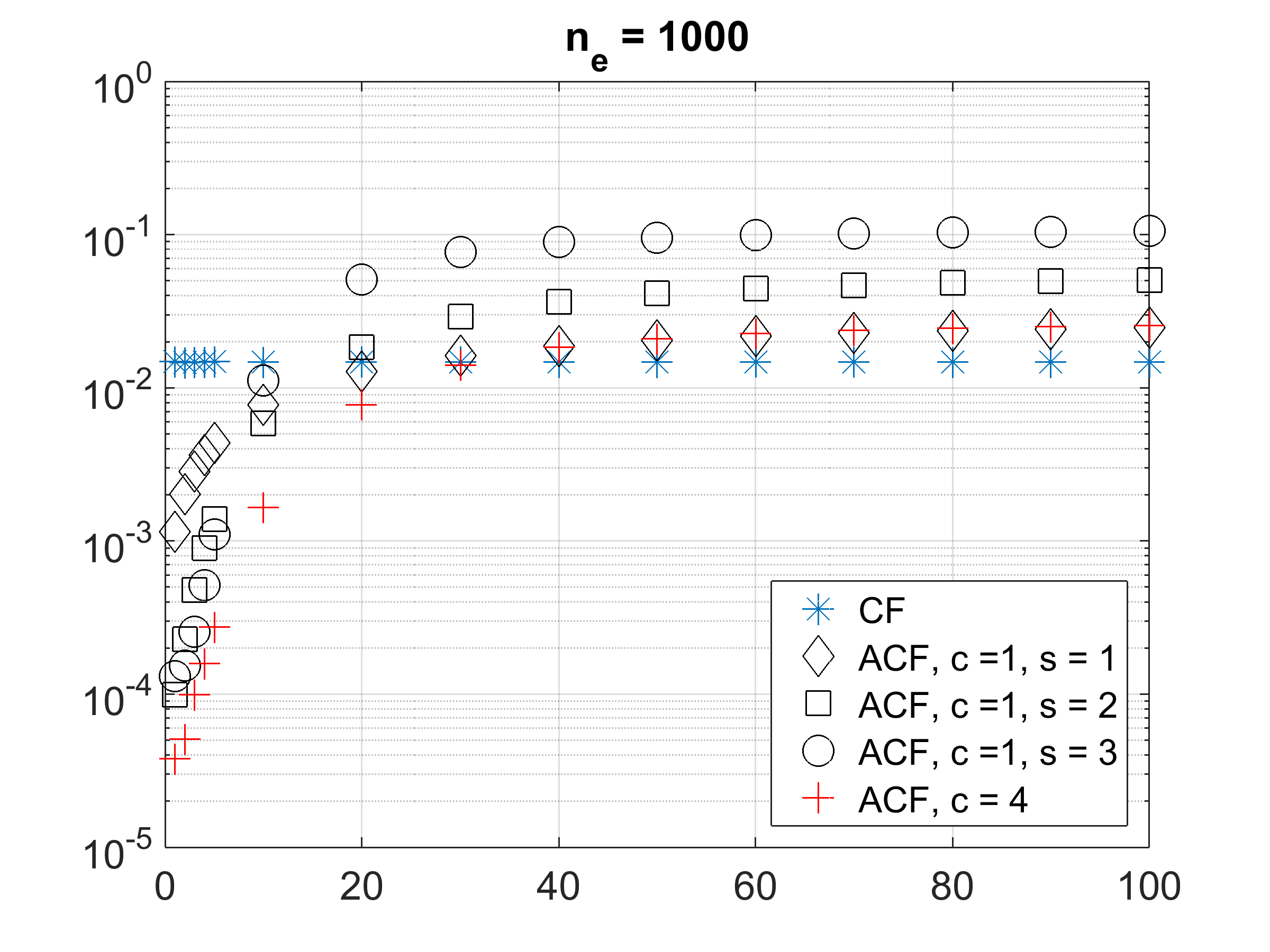}
\caption{False positive rate versus $A/S$ ratio for the different configurations when using 8 bits per cell.}
\label{fig:1}
\end{figure*}

\begin{figure*}[ht]

\includegraphics[width=0.32\textwidth]{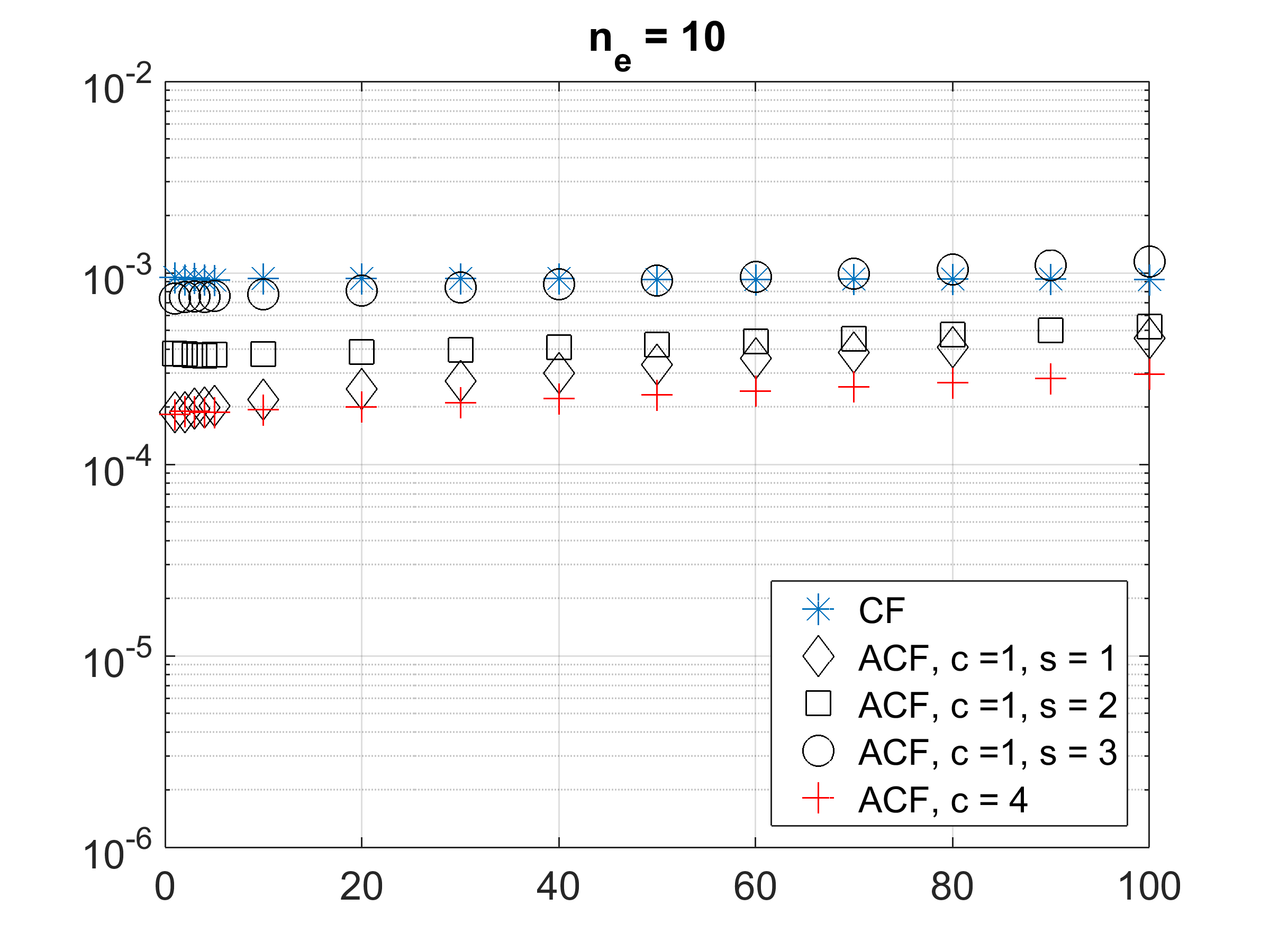}
\includegraphics[width=0.32\textwidth]{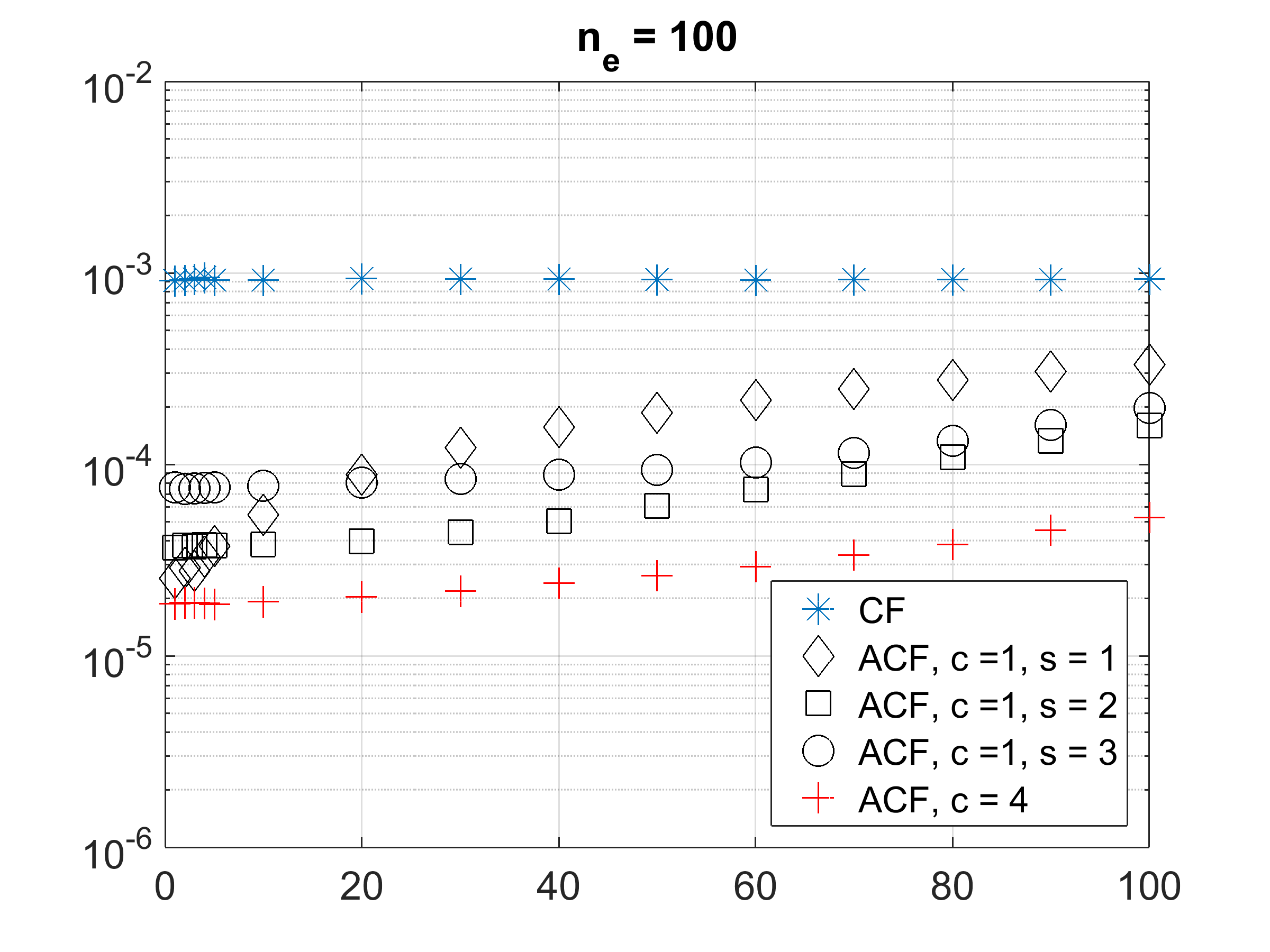}
\includegraphics[width=0.32\textwidth]{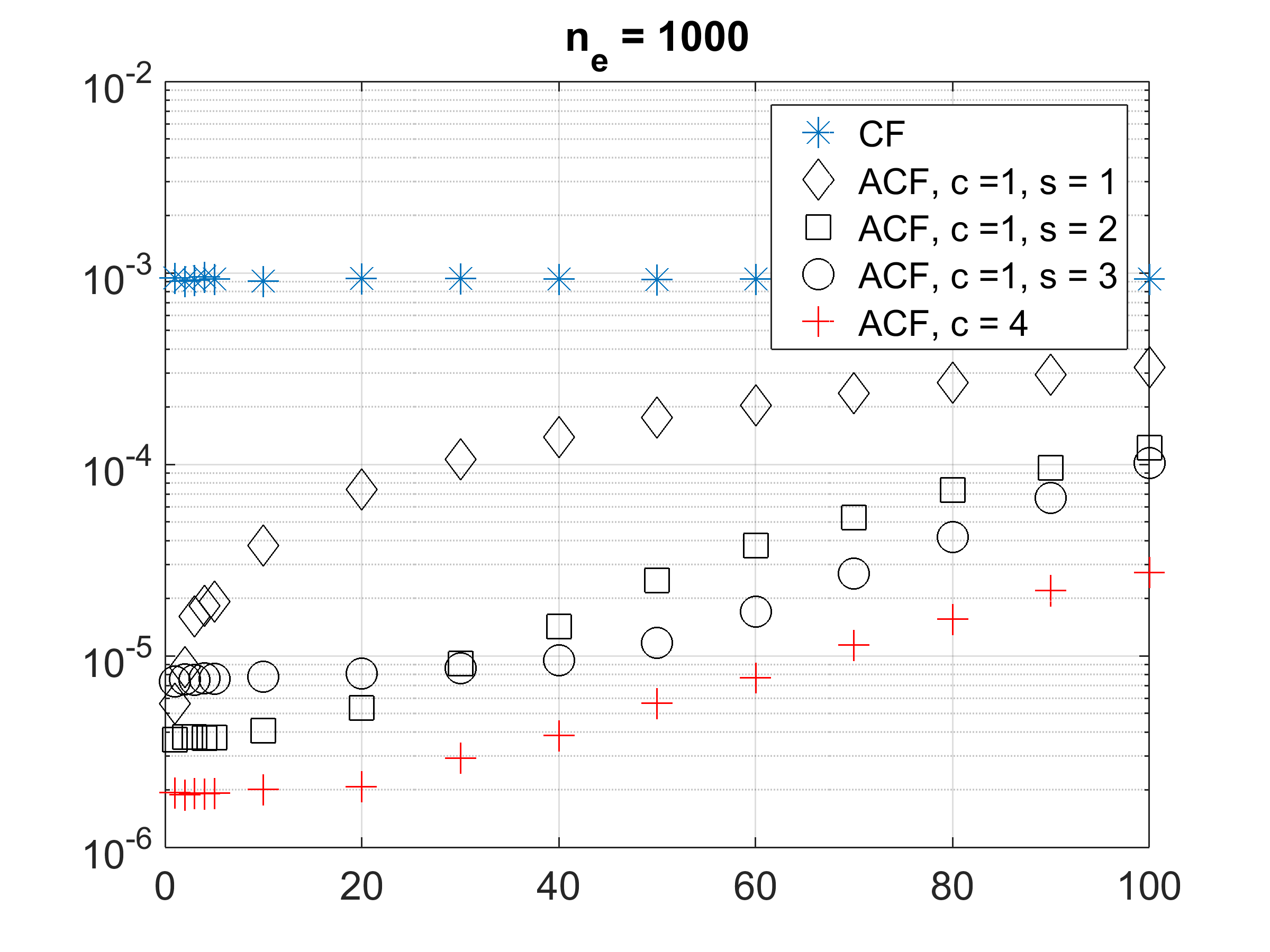}
\caption{False positive rate versus $A/S$ ratio for the different configurations when using 12 bits per cell.}
\label{fig:2}
\end{figure*}

\begin{figure*}[t]

\includegraphics[width=0.32\textwidth]{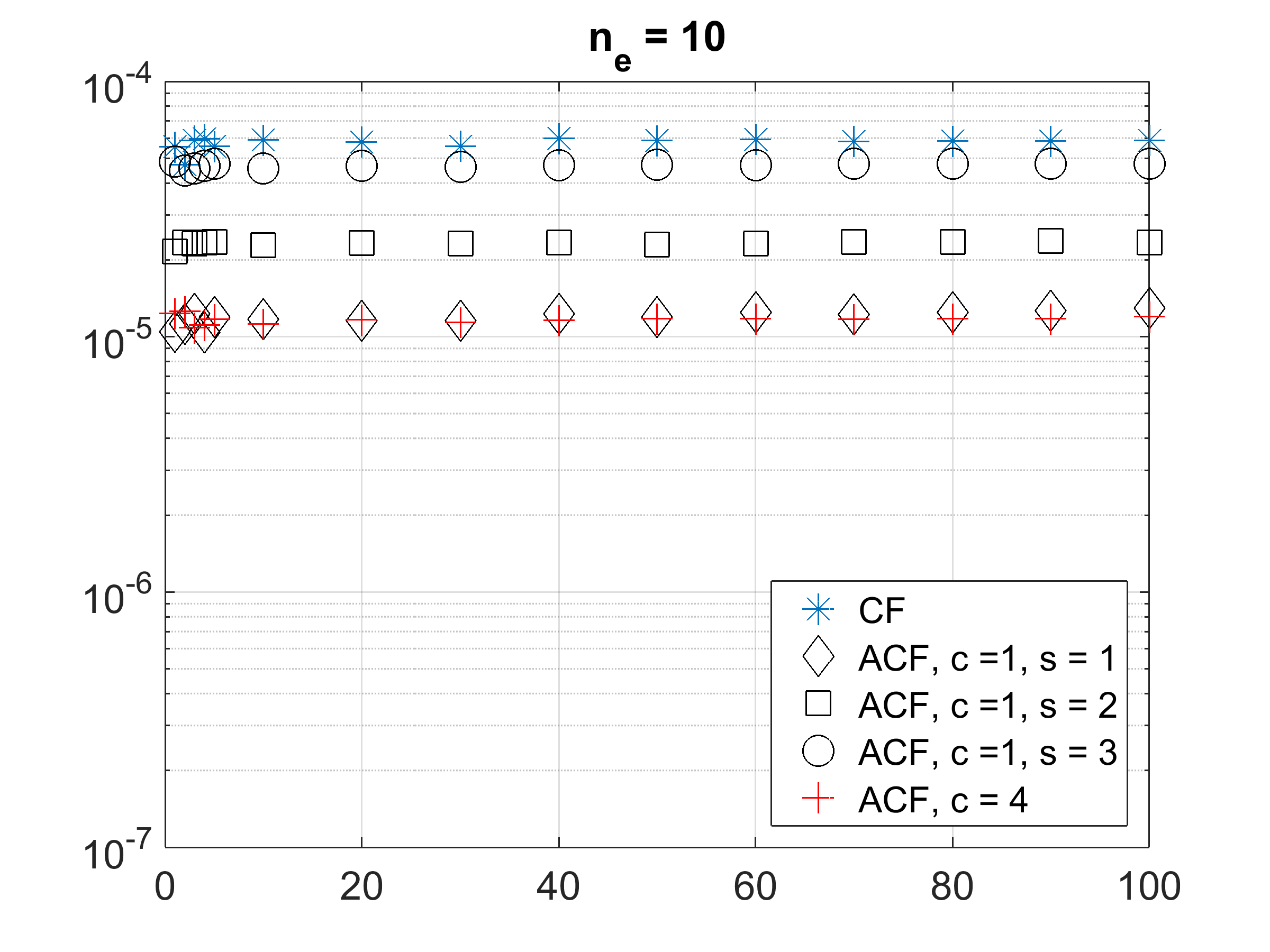}
\includegraphics[width=0.32\textwidth]{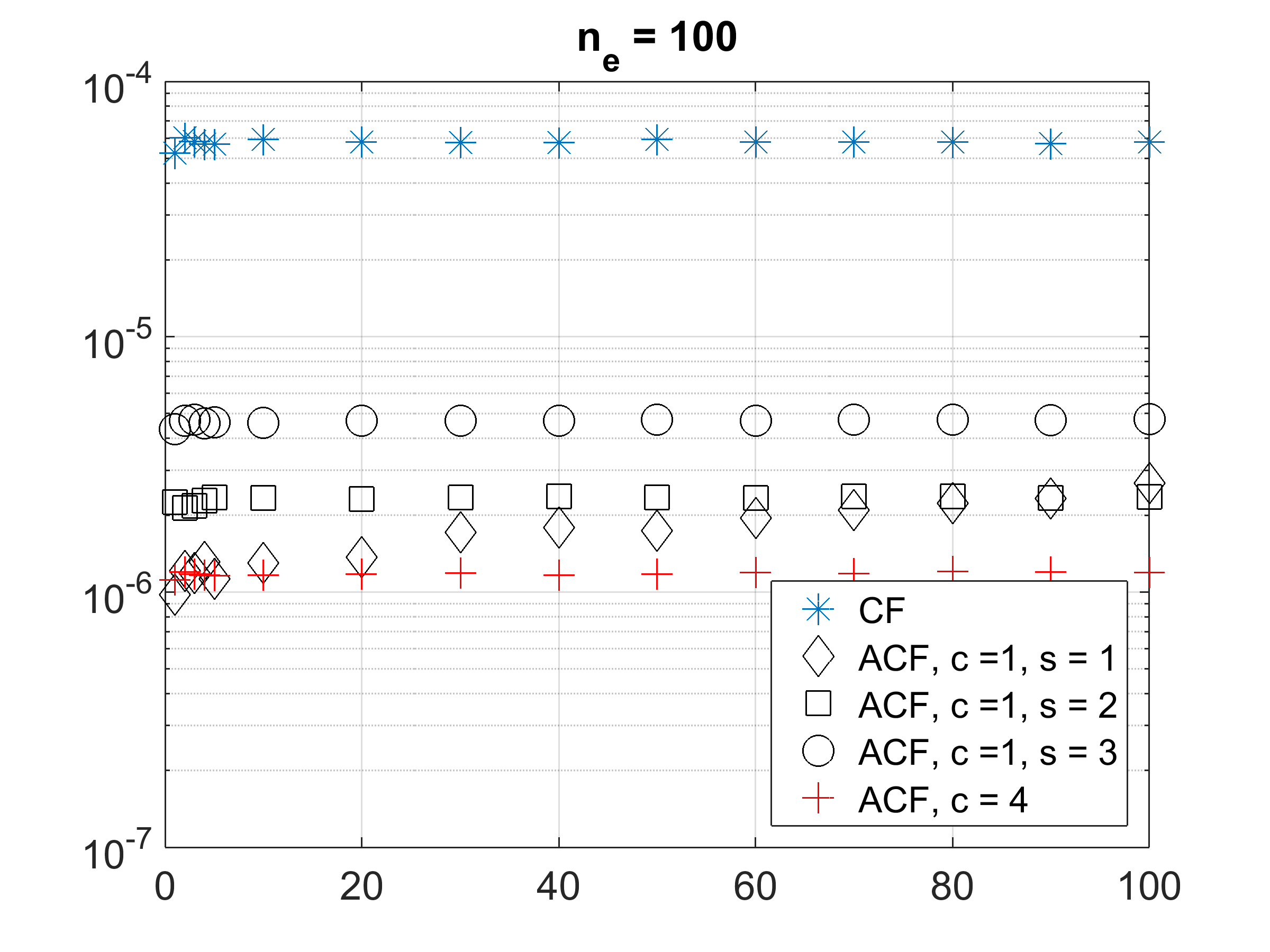}
\includegraphics[width=0.32\textwidth]{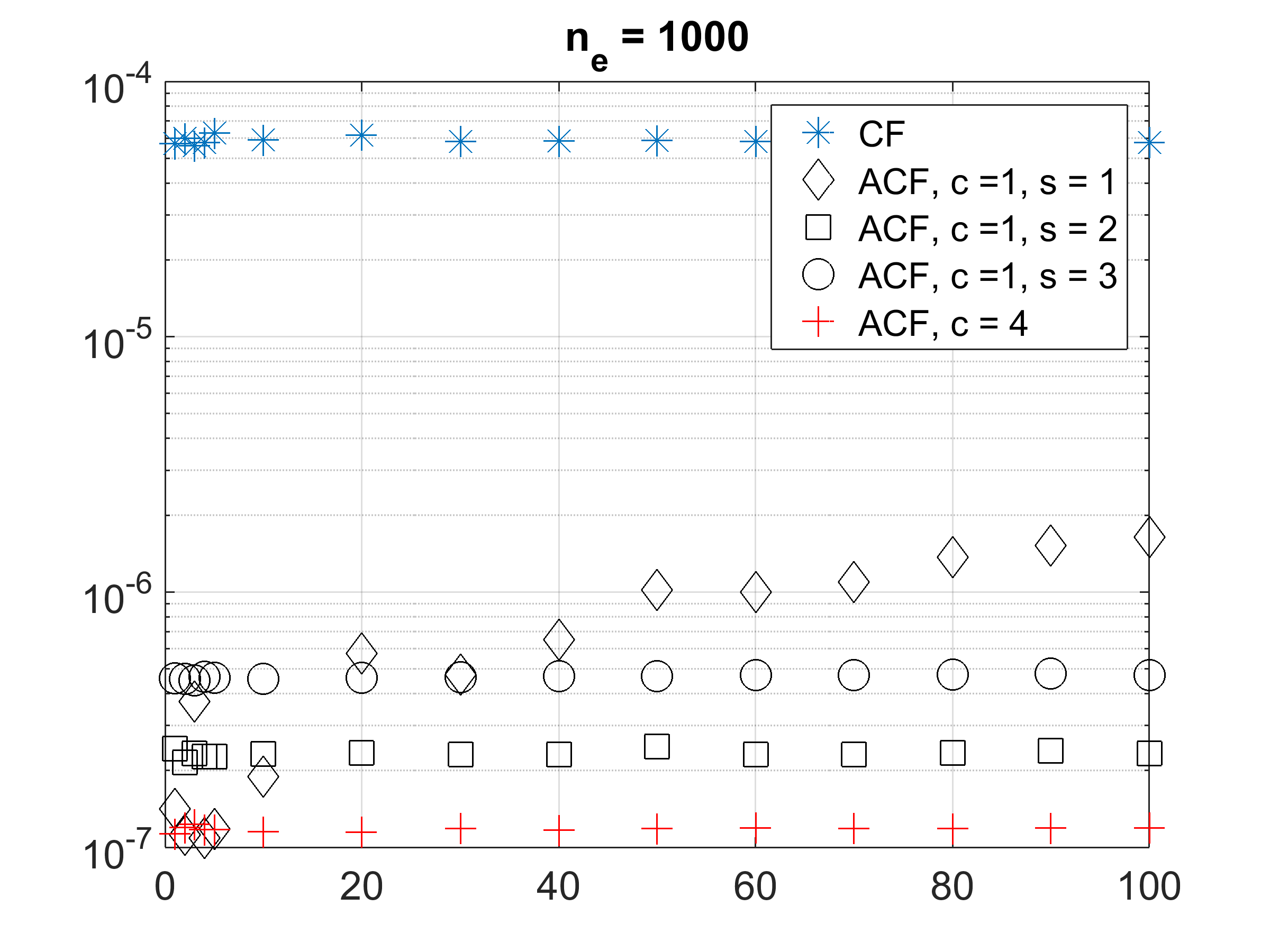}
\caption{False positive rate versus $A/S$ ratio for the different configurations when using 16 bits per cell.}
\label{fig:3}
\end{figure*}

Figures \ref{fig:1}, \ref{fig:2}, and \ref{fig:3} present results for 8, 12 and 16 bits per cell respectively. The average over 10 trials is reported. We observe, as expected, that the false positive rates for the ACFs are better for lower $A/S$ ratios, and that the performance of the ACF improves when the number of queries per element $n_e$ increases. We also see that the second variant of the ACF ($d = 2, c = 4$) provides better performance than the first variant ($d = 4, c = 1$) in almost all cases. This result backs the intuition that the second variant would perform better as it does not require any bits to track the hash function currently selected in the cell.

Focusing on a comparison with a standard cuckoo filter, the ACFs outperform the cuckoo filter for all $A/S$ ratios considered when the number of bits per cell is 12 or 16. (Here we count only the data in the filter part, and do not count the additional memory for the original data.)  When the number of bits per cell is 8, the ACF has a lower false positive rate only when the $A/S$ ratio is small. In configurations where $A/S$ is large and the number of bits per cell is small, there are a large number of false positives, causing many cells to fail to adapt sufficiently, as they simply rotate through multiple false positives.  In contrast, the reductions in the false positive rate for ACFs can be of several orders of magnitude for small values of the $A/S$ ratio and large values of $n_e$. This confirms the potential of the ACF to reduce the false positive rate on networking applications. In fact, the configuration considered in which all elements are queried the same number of times is a worst case for the ACF;  if some elements have more queries than others, the ACF is able to more effectively remove the false positives that occur on those elements.  This holds in simulations, and is seen in the next subsection with real data. (Generally network traffic is highly skewed \cite{ZIPF}.)


Finally, the theoretical estimates were computed using the Markov chains and equations discussed in the previous section for all the configurations.The estimates matched the simulation results and the error was in most cases well below 5\%.  We note that in cases with larger error, the deviations appear to be primarily due to the variance in the simulation results;  low probability events can have a significant effect on the false positive rate. For example, in the first ACF variant when $s=1$ (so there are two fingerprints available for an element in the cell), the probability of having a false positive on the two fingerprints when the $A/S$ ratio is 5 and the number of bits per cell is 16 is below $10^{-6}$, but when that event occurs it can create up to $n_e$ false positives and thereby significantly contribute to the false positive rate, particularly when $n_e$ is large.  Our results demonstrate that the theoretical estimates can be used as an additional approach to guide choosing parameters for performance, along with simulations.

\subsection{Simulations with packet traces}

We also utilized packet traces taken from the CAIDA datasets \cite{CAIDA} to validate the ACF with real-world data. For each trace, a number of flows was selected to be in the set $S$ and placed on the ACF. Then all the packets were queried and the false positive rate was logged. To simulate several values of the $A/S$ ratio, the size of the ACF was changed. The two variants of the ACF and the cuckoo filter were evaluated as in the previous subsection. The ACF was able to consistently reduce the false positive rate of the cuckoo filter.
We selected the 5-tuple as the key to insert in the ACF, and the set of 5-tuples in each trace as the $A \cup S$ set. Then we select the same $A/S$ ratios used in the previous experiment and set the size of the $S$ set, and consequently the size of the ACF so that the it reaches the 95\% load when it is filled. After, the ACF is loaded picking the first $|S|$ 5-tuples in the trace. The remaining flows were considered as potential false positives for the ACF. 


We provide results for three distinct traces\footnote{The traces used were taken from the CAIDA 2014 dataset and are equinix-chicago.dirA.20140619-130900, equinix-chicago.dirB.20140619-132600, and equinix-sanjose.dirA.20140320-130400.}, with each corresponding to 60 seconds of traffic. The first trace has approximately 18.5 million packets and 691,371 different flows, giving an average value of $n_e \approx 26.7$ packets per flow. The distribution of $n_e$ is highly skewed with the largest flow having 130,675 packets, the largest one thousand flows each having more than ten thousand packets and the largest ten thousand flows each having more than two thousand packets. The other two traces have 14.6 and 37.4 million packets and 632,543 and 2,313,092 flows respectively, and the number of packets per flow is also highly skewed.

\begin{figure*}[t]
	\includegraphics[width=0.32\textwidth]{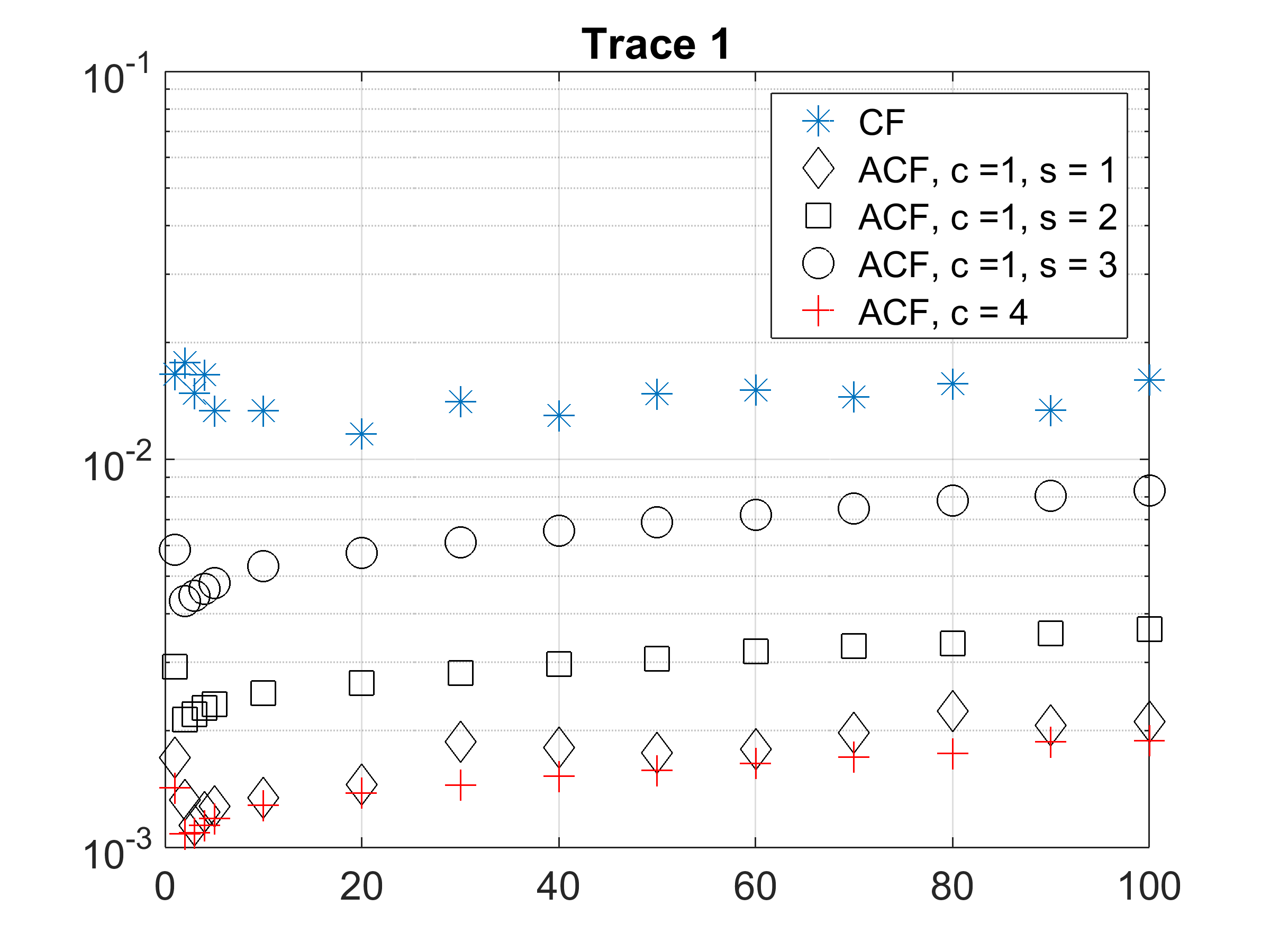}
    \includegraphics[width=0.32\textwidth]{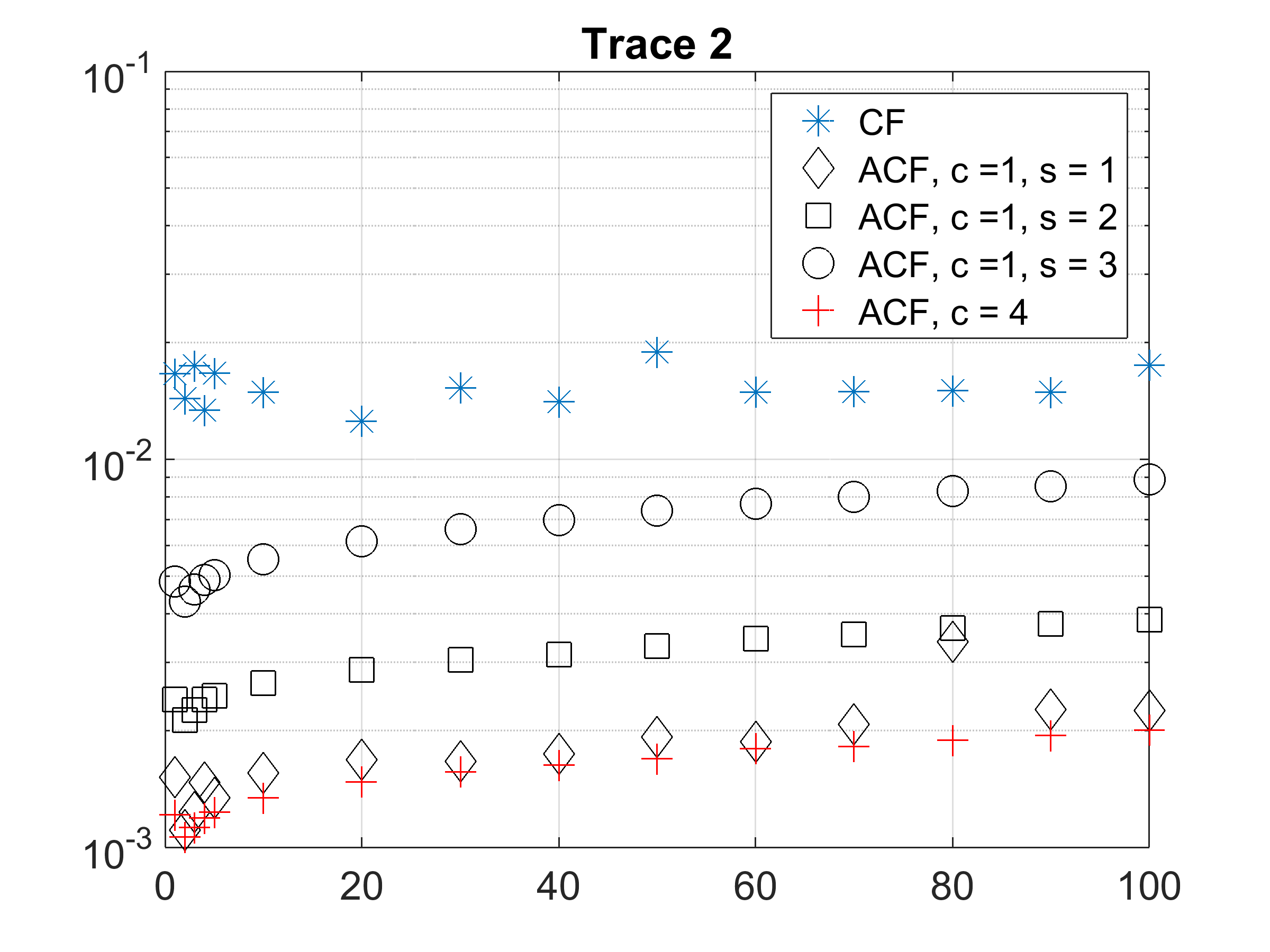}
    \includegraphics[width=0.32\textwidth]{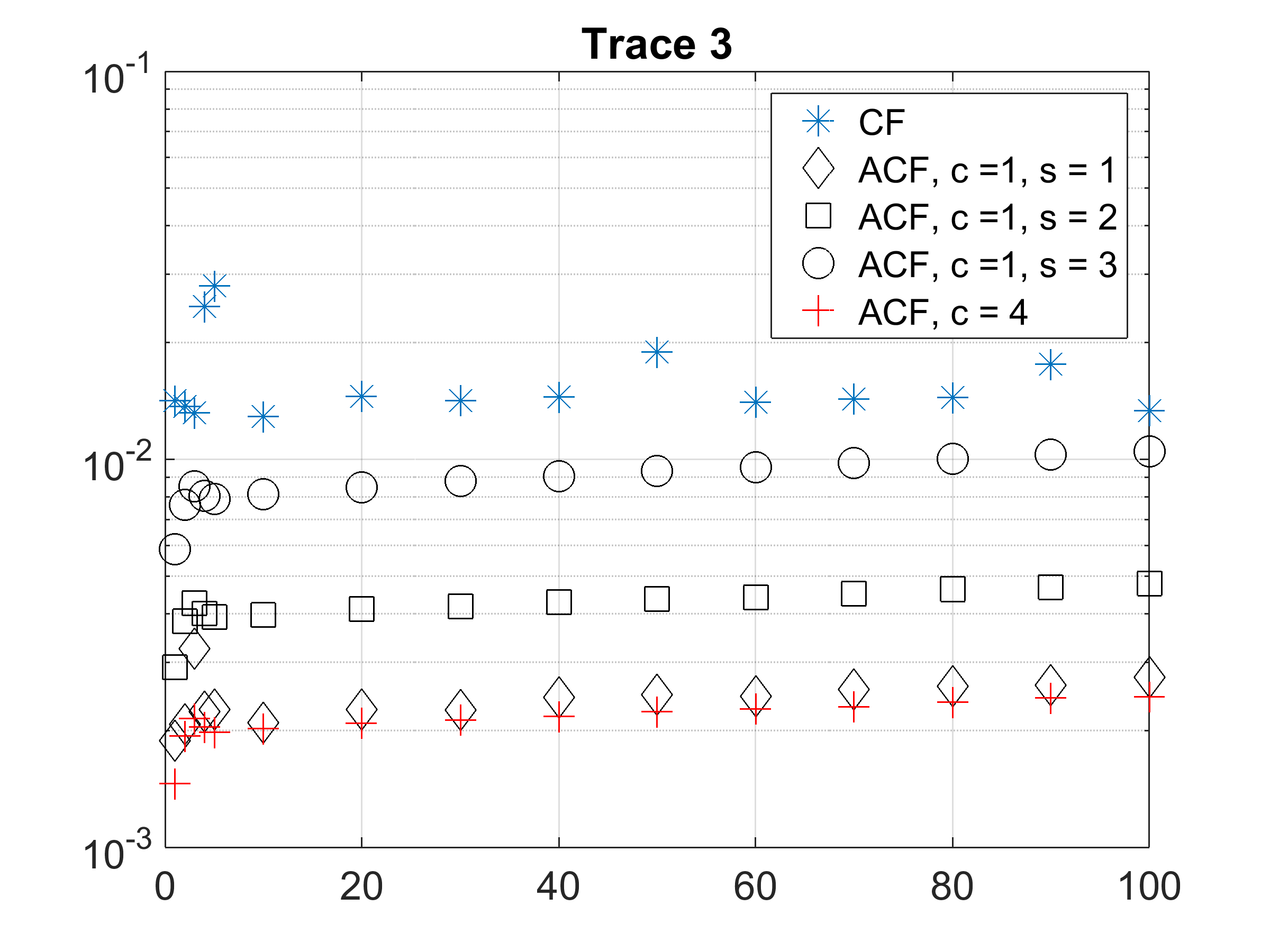}
	\caption{False positive rate versus $A/S$ ratio for the three CAIDA traces when using 8 bits per cell.}
	\label{fig:4}
\end{figure*}

\begin{figure*}[t]
	\includegraphics[width=0.32\textwidth]{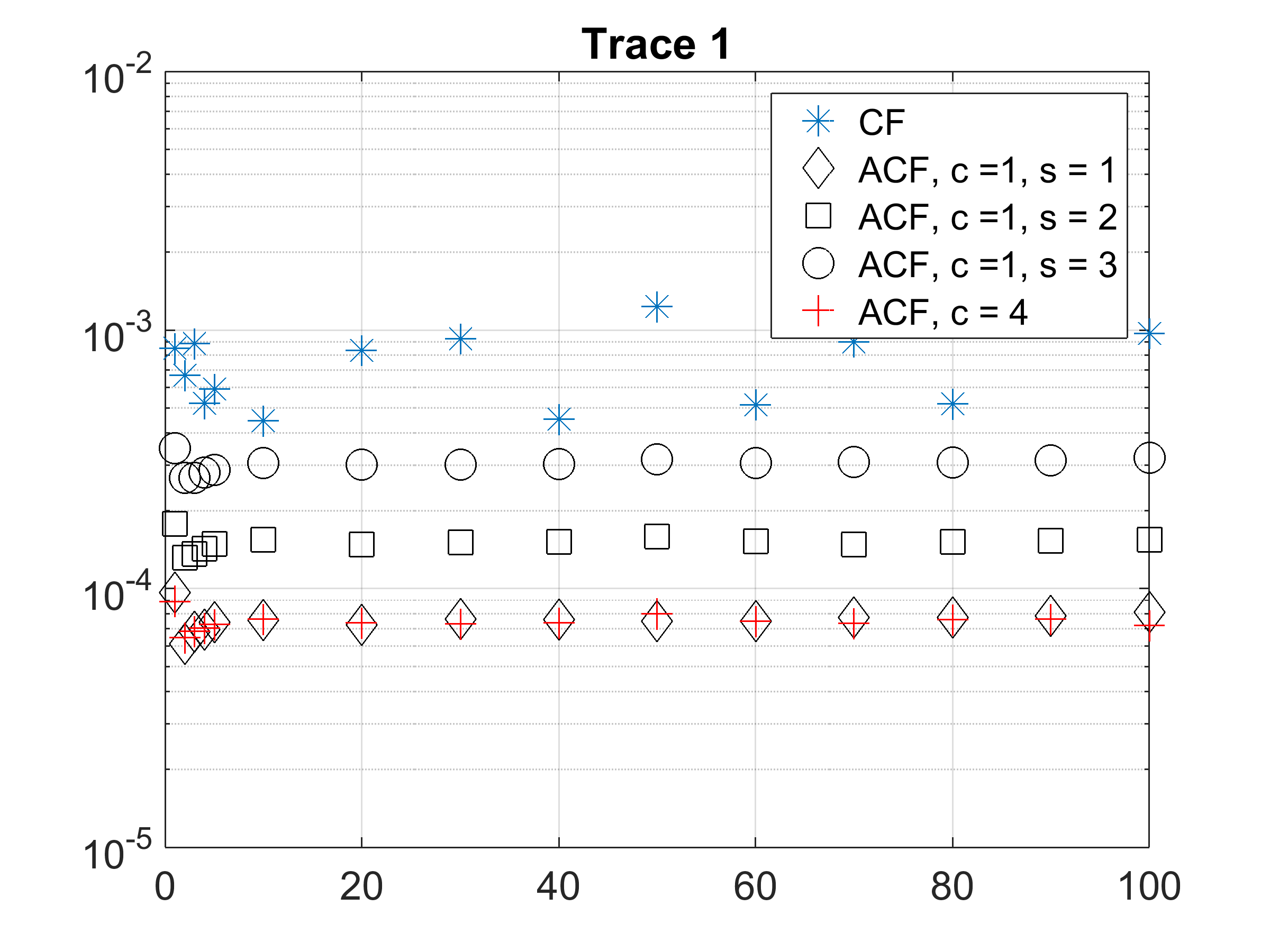}
    \includegraphics[width=0.32\textwidth]{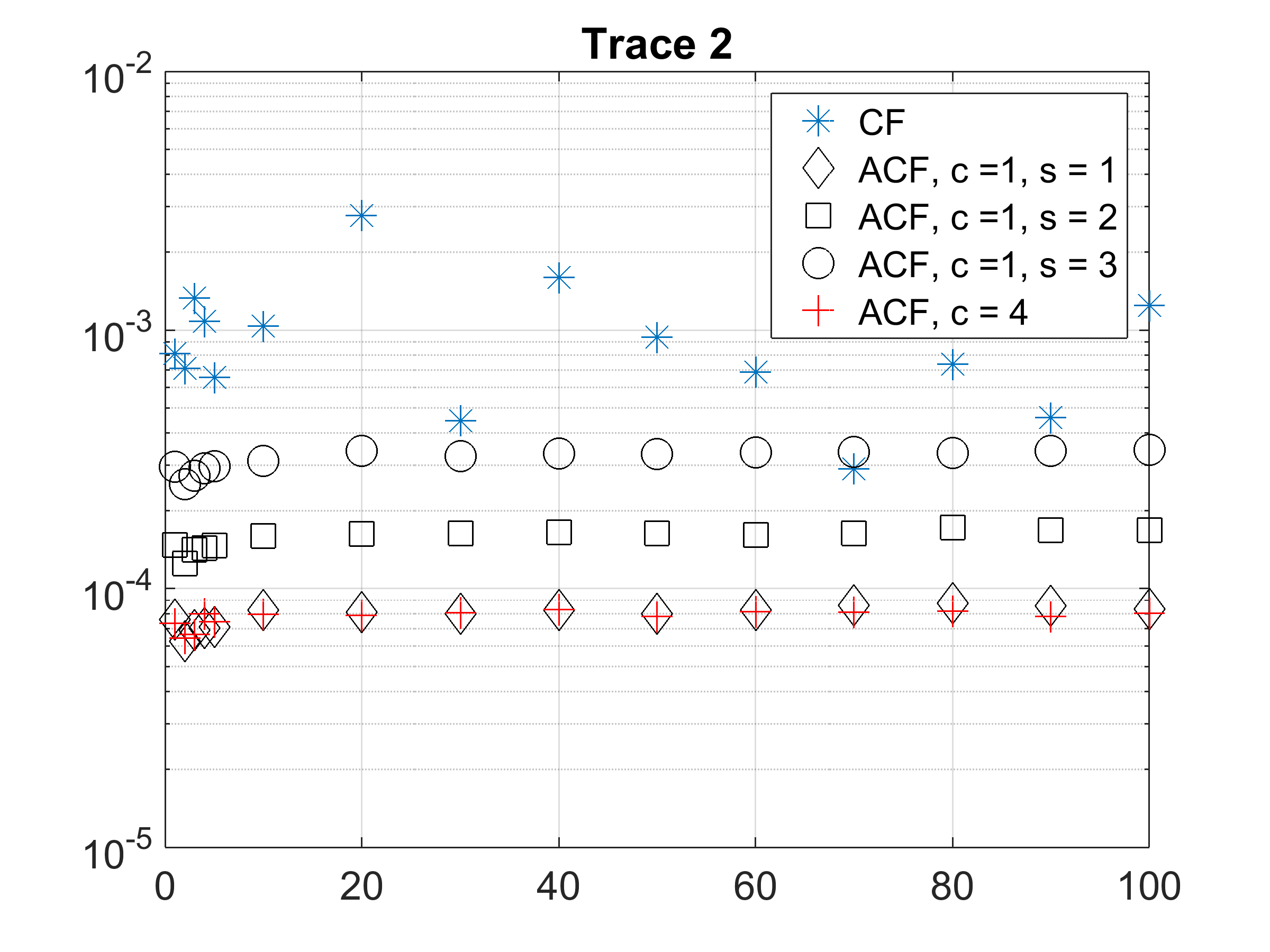}
    \includegraphics[width=0.32\textwidth]{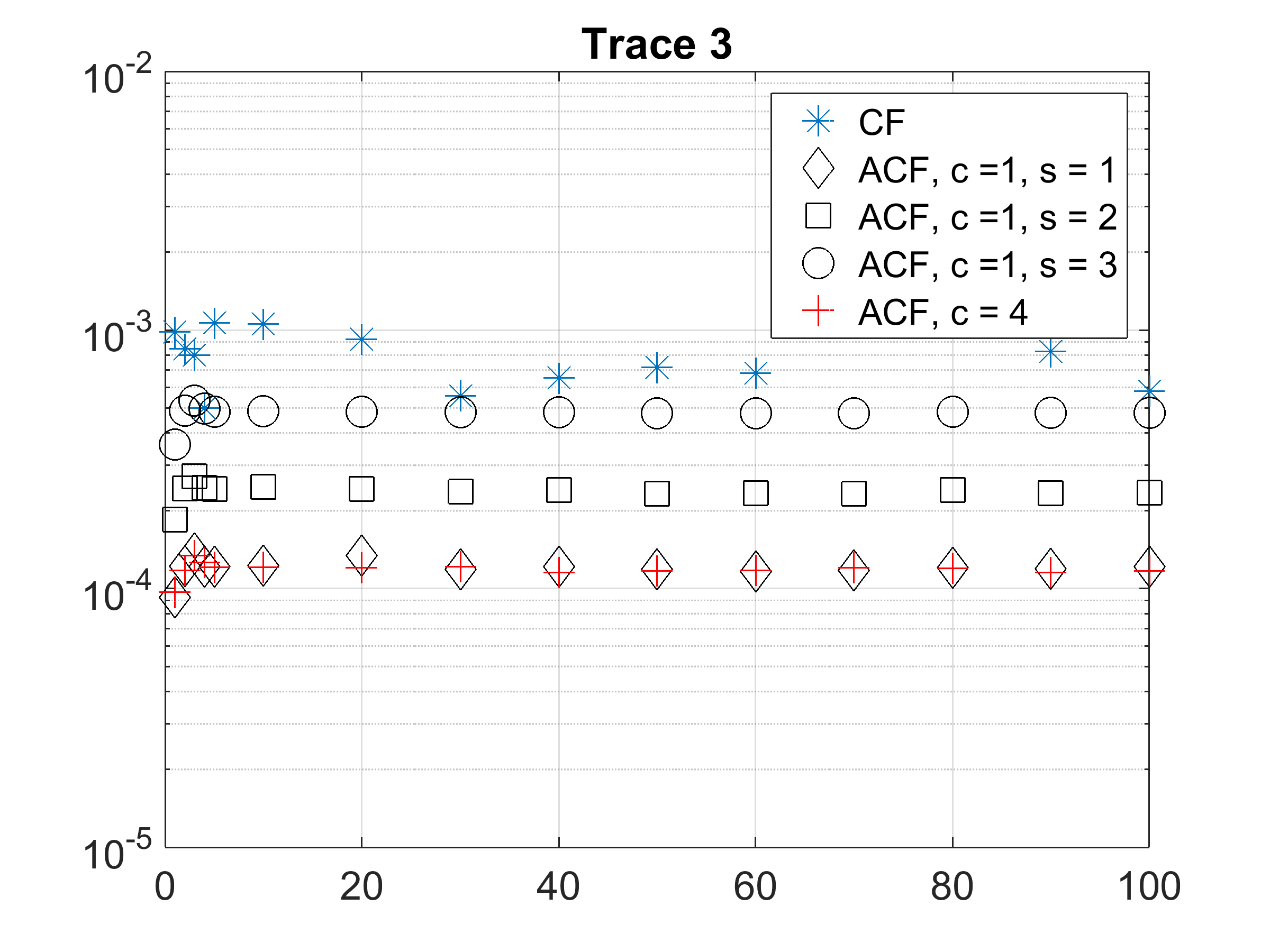}
	\caption{False positive rate versus $A/S$ ratio for the three CAIDA traces when using 12 bits per cell.}
	\label{fig:5}
\end{figure*}

\begin{figure*}[t]
	\includegraphics[width=0.32\textwidth]{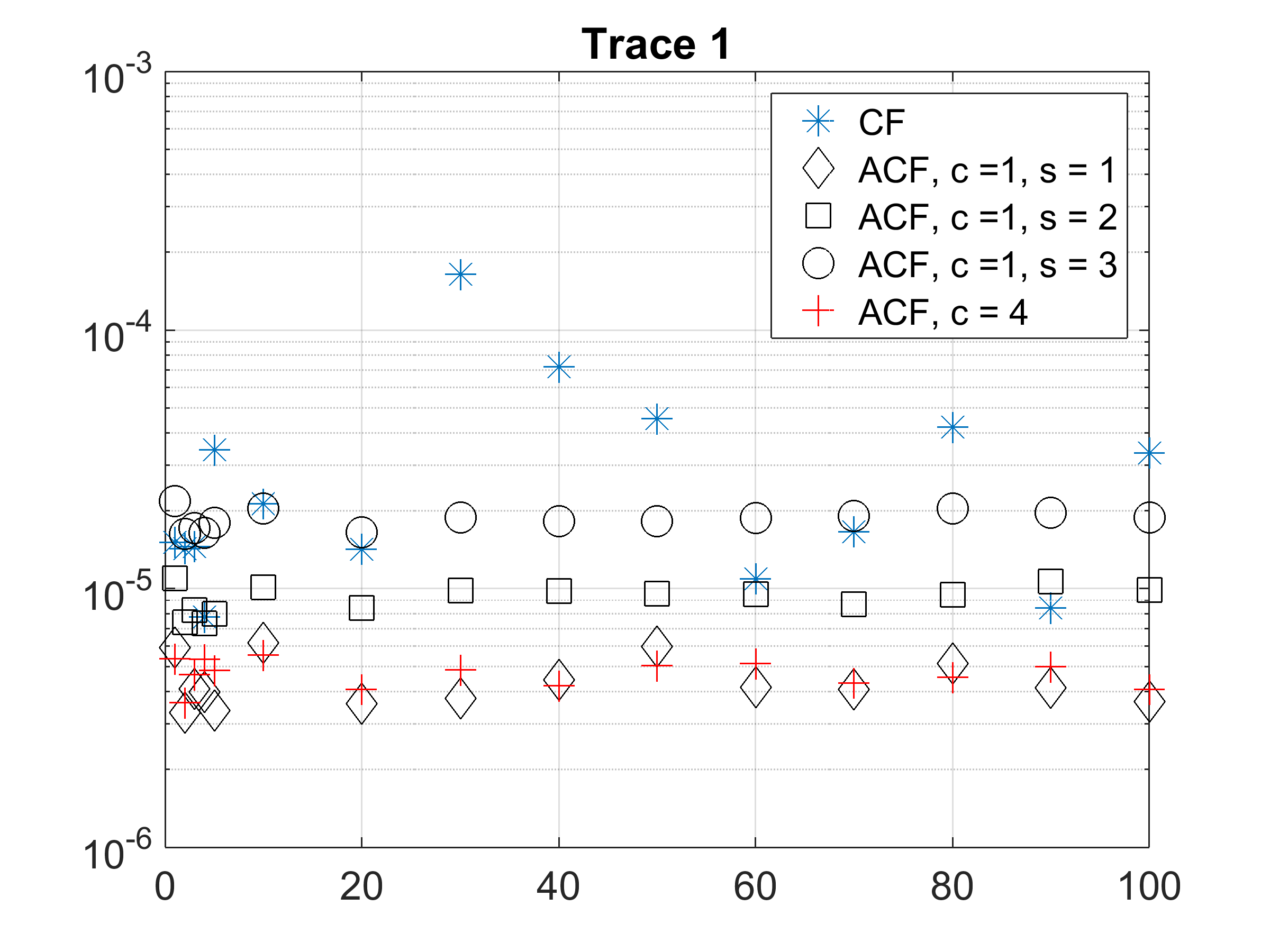}
    \includegraphics[width=0.32\textwidth]{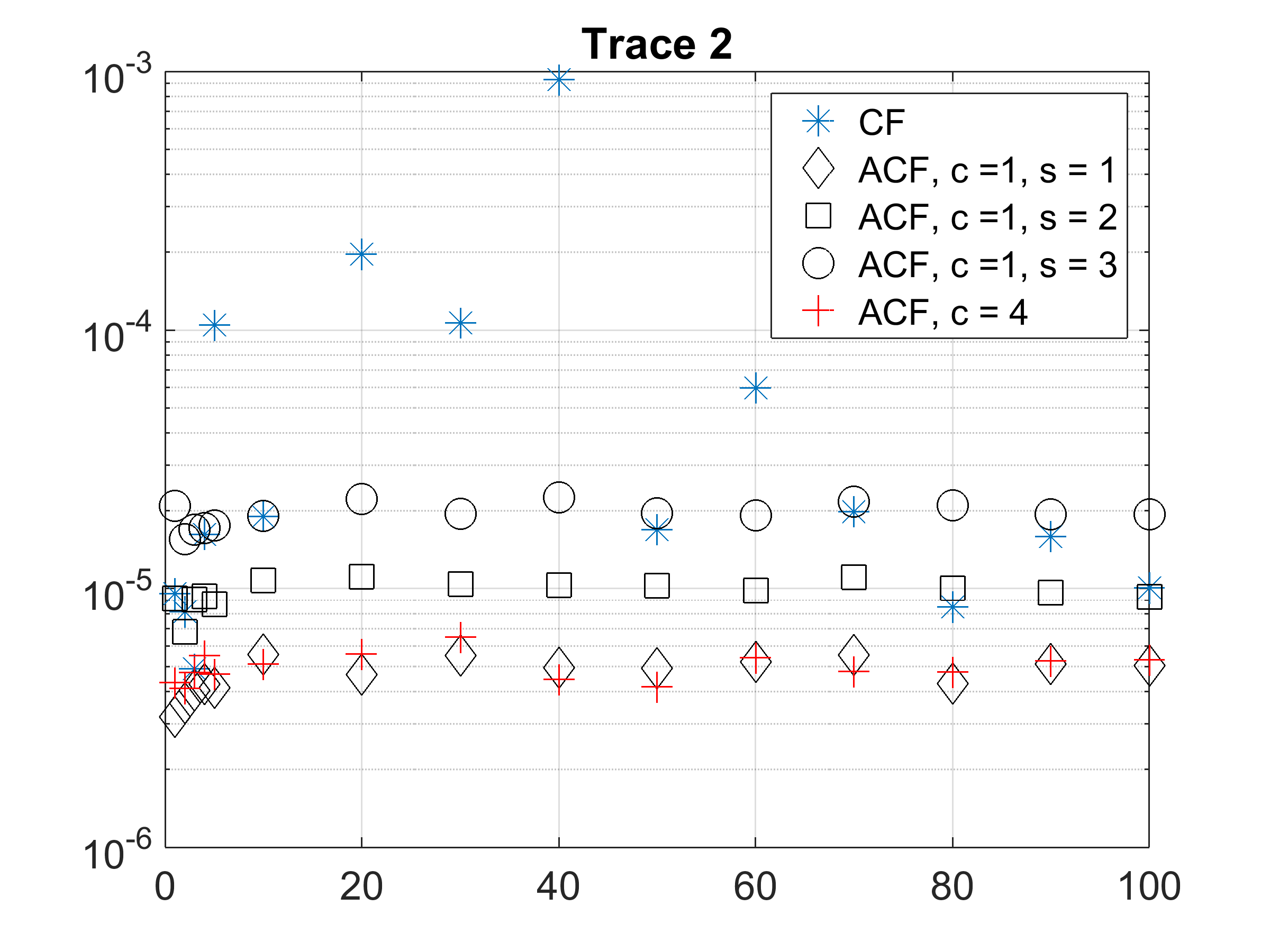}
    \includegraphics[width=0.32\textwidth]{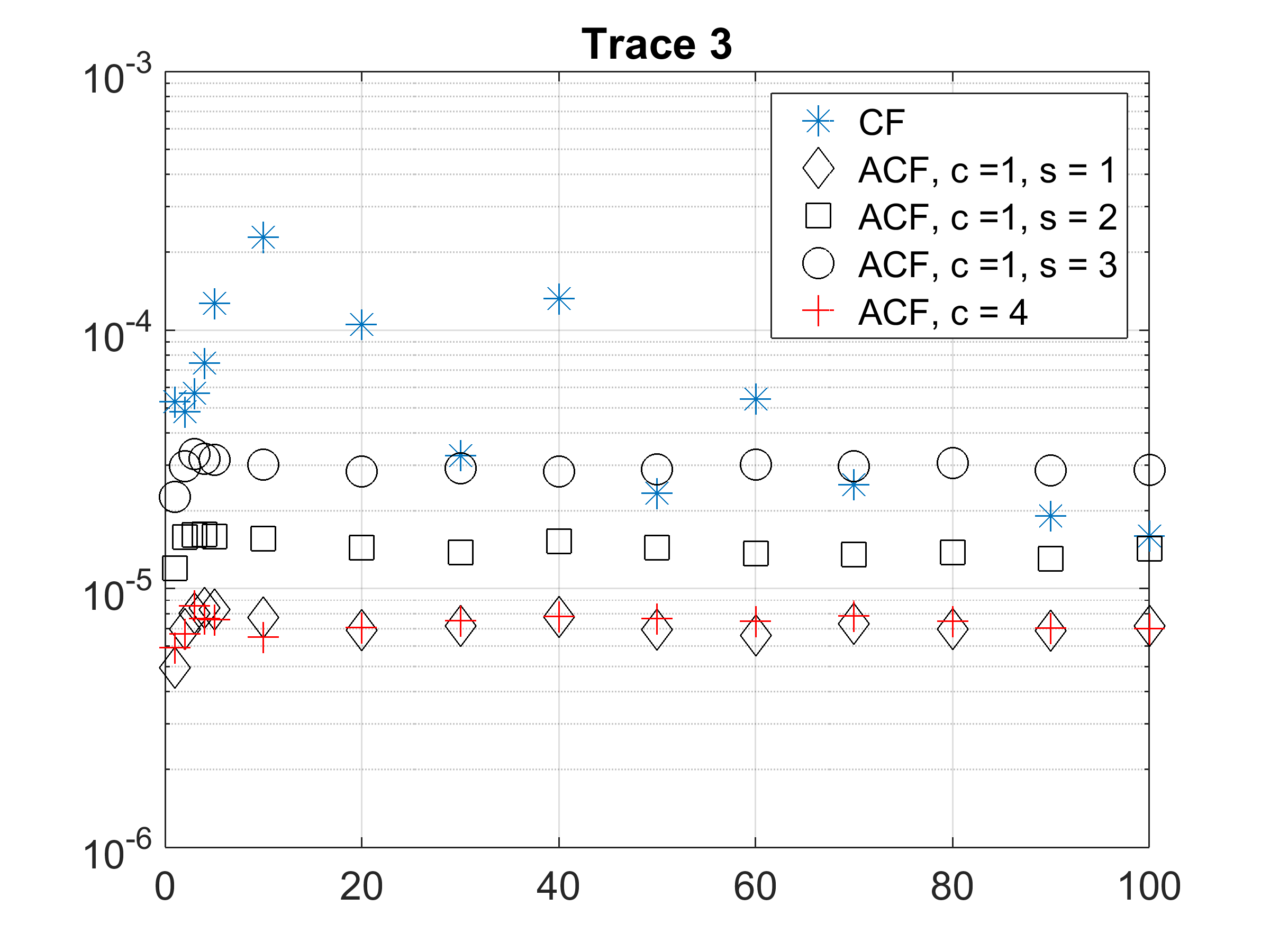}
	\caption{False positive rate versus $A/S$ ratio for the three CAIDA traces when using 16 bits per cell.}
	\label{fig:6}
\end{figure*}

The false positive rates for the three traces are shown in Figures \ref{fig:4}, \ref{fig:5} and \ref{fig:6} for 8, 12, and 16 bits per cell. The results are similar for the three traces. The best ACF configurations are the first variant ($d = 2, c = 1$) with $s = 1$ and the second variant ($d = 2, c = 4$). Both provide similar false positive rates and clearly outperform the cuckoo filter for all the values of the bits per cell tested. An interesting observation is that the ACF reduces the false positive rate for all $A/S$ ratios even when the number of bits per cell is only eight. This was not the case with simulated queries, and the empirical result can be explained by how the ACF benefits from the skewness of the traffic. For example, let us consider the first ACF first variant ($d = 2, c = 1$) with $s = 1$, and consider a case where there are two flows that create false positives on a bucket, one on each of the hash selector values. If one of the flows has 1000 packets and the other has only 10, in the worst case, the ACF gives only 20 false positives on the bucket, because after 10 packets the hash selector value that gave a false positive for the small flow will give no more false positives, and the ACF will stop on that hash selector.  Also, when the number of bits per cell increases, the results for the cuckoo filter show high variability. This is again due to the traffic skewness because as the false positive probability gets smaller, only a few flows contribute to the false positive rate, and the rate therefore depends heavily on the number of packets of those flows. This variability does not appear on the ACF as false positives on flows with many packets are effectively removed by the ACF as explained before.

\section{Conclusions and future work}

This paper has presented the adaptive cuckoo filter (ACF), a variant of the cuckoo filter that attempts to remove false positives after they occur so that the following queries to the same element do not cause further false positives, thereby reducing the false positive rate. The performance of the ACF has been studied theoretically and evaluated with simulations. The results confirm that when queries have a temporal correlation the ACF is able to significantly reduce the false positive rate compared to the cuckoo filter, which itself offers improvements over the original Bloom filter. 

The main novelty of the ACF is its ability to adapt to false positives by identifying the element that caused the false positive, removing it and re-inserting it again in a different way so that the false positive is removed, but the element can still be found when searched for. This paper has studied two variants of the ACF, in both of which the elements remain in the same bucket when false positives are found; other more complex schemes could move elements to different buckets as well, but we leave studying such variants to future work.  We believe the simple variants we have proposed, because of their simplicity and strong performance, can find uses in many natural applications.  

\section*{Acknowledgments}
Michael Mitzenmacher is supported in part by NSF grants CCF-1563710, CCF-1535795, CCF-1320231, and CNS-1228598. 
Salvatore Pontarelli is partially supported by the Horizon 2020 project 5G-PICTURE (grant \#762057). 
Pedro Reviriego would like to acknowledge the support of the excellence network Elastic Networks TEC2015-71932-REDT funded by the Spanish Ministry of Economy and competitivity.


\onecolumn
\section*{Appendix: Algorithms}
We provide pseudocode for our algorithms.  

\begin{algorithm}
\caption{ACF for buckets with a single cell: Lookup for an element $x$} \label{Algo1}
\begin{algorithmic}[1]
\Require Element $x$ to search for
\Ensure Positive/negative
\For{$i\leftarrow 1$ to $4$}
  \State Access bucket $h_i(x)$
  \State Read $\alpha$ and $f_{\alpha}(y)$ stored on the cell
  \State Compute $f_{\alpha}(x)$
  \State Compare $f_{\alpha}(y)$ in the cell with $f_{\alpha}(x)$
  \If{Match}
	\State \textbf{return} positive
  \EndIf         
\EndFor
  \State \textbf{return} negative 

\end{algorithmic}
\end{algorithm}
\vspace{-8pt}

\begin{algorithm}
\caption{ACF for buckets with a single cell: Adaptation to remove a false positive} \label{Algo2}
\begin{algorithmic}[1]
\Require False positive on bucket $h_i(y)$
\Ensure update $\alpha$ and $f_{\alpha}(y)$ 
\State  Access bucket $h_i(y)$\; 
\State  Read $\alpha$ stored on the cell\;
\State  Retrieve element $y$ stored on bucket $h_i(y)$ from the main table\;
\State  Increment $\alpha$ modulo $2^s$\;
\State  Compute $f_{\alpha}(y)$\;
\State  Store the new value of $\alpha$ and $f_{\alpha}(y)$ on bucket $h_i(y)$ in the ACF\;

\end{algorithmic}
\end{algorithm}

\vspace{-8pt}
\begin{algorithm}
\caption{ACF for buckets with multiple cells: Lookup for an element $x$} \label{Algo3}
\begin{algorithmic}[1]
\Require Element $x$ to search for
\Ensure Positive/negative 
\State Compute $f_1(x),f_2(x),f_3(x),f_4(x)$\;
\For{$i\leftarrow 1$ to $2$}
\State    Access bucket $h_i(x)$
    \For{$j\leftarrow 1$ to $4$}
\State    Compare fingerprint stored in cell $j$ with $f_j(x)$\; 
    \If{Match}
\State \textbf{return} positive
\EndIf         
\EndFor
\EndFor
\State \textbf{return} negative

\end{algorithmic}
\end{algorithm}

\vspace{-8pt}
\begin{algorithm}[ht!]
\caption{ACF for buckets with multiple cells: Adaptation to remove a false positive} \label{Algo4}
\begin{algorithmic}[1]
\Require False positive in cell $j$ on bucket $h_i(y)$
\Ensure update bucket $h_i(y)$
\State Select randomly a cell $k$ different from $j$ on bucket $h_i(y)$\;
\State Retrieve elements $y,z$  stored in cells $j,k$ on bucket $h_i(y)$ from the main table\;
\State Compute $f_j(z)$ and $f_k(y)$\;
\State Write $f_j(z)$ and $f_k(y)$ in cells $j$ and $k$ on bucket $h_i(y)$ of the ACF\;
\State Write $z$ in cell $j$ and $y$ in cell $k$ on  bucket $h_i(y)$ in the main table\;

\end{algorithmic}
\end{algorithm}

\end{document}